\documentclass[12pt]{article}
\usepackage{amsmath,amssymb,amsthm,mathtools}
\usepackage{natbib}
\usepackage{graphicx}
\usepackage[nodisplayskipstretch]{setspace} %
\usepackage[inline]{enumitem}
\usepackage{fullpage,rotating,afterpage}
\usepackage{microtype} %

\usepackage[hidelinks]{hyperref}
\usepackage{pdflscape}
\usepackage{float}

\usepackage[T1]{fontenc}

\usepackage{tikz}
\usepackage{pdflscape}
\usepackage{booktabs}
\usepackage[justification=justified]{caption}
\usepackage[flushleft]{threeparttable}
\usepackage{multirow}

\newcommand{\E}{\mathbb{E}}
\newcommand{\V}{\mathbb{V}}
\newcommand{\IR}{\mathbb{R}}

\newcommand{\1 }[1]{\mathbf{1}\{#1\}}
\newcommand{\Var}{\textrm{Var}}

\newcommand{\Cov}{\textrm{Cov}}

\newcommand{\sgn}{\textnormal{sign}}
\newcommand{\wh}{\widehat}

\newtheorem{proposition}{Proposition}

\numberwithin{equation}{section}

\usepackage{verbatim,color}

\def\boxit#1{\vbox{\hrule\hbox{\vrule\kern6pt
\vbox{\kern6pt#1\kern6pt}\kern6pt\vrule}\hrule}}

\DeclareMathOperator*{\argmin}{argmin}

\newcommand{\captionfonts}{\small}

\makeatletter
\long\def\@makecaption#1#2{%
\vskip\abovecaptionskip
\sbox\@tempboxa{{\captionfonts #1: #2}}%
\ifdim \wd\@tempboxa >\hsize
	{\captionfonts #1: #2\par}
\else
	\hbox to\hsize{\hfil\box\@tempboxa\hfil}%
\fi
\vskip\belowcaptionskip}
\makeatother  

\usepackage{titlesec}
\titleformat{\section}[block]{\centering\normalfont}{\thesection.}{0.5em}{\uppercase }
\titleformat{\subsection}[runin]{\normalfont}{\thesubsection.}{0.4em plus .1em minus .2em}{\bfseries}[.]
\titleformat{\subsubsection}[runin]{\normalfont}{\thesubsubsection.}{0.4em plus .1em minus .2em}{\it}[.]
\titlespacing*\section{0pt}{18pt plus 4pt minus 2pt}{5pt plus 2pt minus 2pt}
\titlespacing*\subsection{0pt}{10pt plus 2pt minus 1pt}{5pt plus 2pt minus 2pt }
\titlespacing*\subsubsection{0pt}{4pt plus 1pt minus 1pt}{5pt plus 2pt minus 2pt}

\usepackage{chngcntr}
\usepackage{apptools}
\AtAppendix{\counterwithin{lemma}{section}}
\AtAppendix{\counterwithin{assumption}{section}}
\AtAppendix{\counterwithin{corollary}{section}}
\AtAppendix{\counterwithin{theorem}{section}}
\AtAppendix{\counterwithin{proposition}{section}}
\AtAppendix{\counterwithin{table}{section}}
\AtAppendix{\counterwithin{figure}{section}}
\AtAppendix{\counterwithin{remark}{section}}

\makeatletter
\def\mythanks#1{%
\protected@xdef \@thanks {\@thanks \protect \footnotetext [\the \c@footnote ]{#1}}%
}
\makeatother

\title{\bfseries \large\uppercase{Donut Regression Discontinuity Designs}\mythanks{First version: August 27, 2023. This version: \today. We would like to thank Guido Imbens, Fran\c{c}ois Gerard, and numerous seminar and conference participants for  their helpful comments and suggestions. The authors gratefully acknowledge financial support by the European Research Council (ERC) through grant SH1-77202. 
Claudia Noack, Department of Economics, University of Bonn. E-mail: claudia.noack@uni-bonn.de. Website: https://claudianoack.github.io.
Christoph Rothe, Department of Economics, University of Mannheim, 68131 Mannheim, Germany.
E-mail: rothe@vwl.uni-mannheim.de. Website: www.christophrothe.net. }}

\author{\textsc{Claudia Noack} \and \textsc{Christoph Rothe}}
\date{}

\begin{document}

\pagestyle{plain}

\newtheorem{theorem}{Theorem}
\newtheorem{definition}{Definition}
\newtheorem{lemma}{Lemma}
\newtheorem{assumption}{Assumption}
\theoremstyle{definition}
\newtheorem{example}{Example}
\newtheorem{remark}{Remark}

\bibliographystyle{ecta}

\maketitle

\begin{abstract}

	Donut regression discontinuity (RD) designs exclude observations in a neighborhood around the cutoff. They are widely used in empirical work to address concerns about manipulation, sorting, or measurement error in the running variable. Despite their popularity, donut RD estimators are typically implemented heuristically, with limited formal guidance and unclear implications for bias, variance, and statistical inference.
	This paper provides a theoretical analysis of donut RD designs within the local linear estimation framework. We derive new results for point estimation, inference, and specification testing in donut RD designs. We show that
	donut trimming affects the bias and variance of the estimator  and thus introduces substantial additional uncertainty relative to conventional RD estimators. We further show that recently developed bias-aware confidence intervals accurately account for this additional uncertainty without requiring modifications. Finally, we formalize specification tests based on comparing conventional and donut RD estimators to allow for valid statistical inference. The results are illustrated with two empirical applications.
\end{abstract}

\onehalfspacing

\section{Introduction}
Regression discontinuity (RD) designs provide a framework for identifying causal effects from observational data under clear and interpretable assumptions. In empirical applications in economics and the social sciences, estimation and inference commonly rely on local linear regression, whose statistical properties have been extensively studied \citep[e.g.,][]{hahn2001identification, imbens2012optimal, calonico2014robust, armstrong2018optimal}. In this paper, we study a widely used variant of the standard RD design that excludes observations near the treatment threshold. This approach is commonly referred to as a ``donut'' RD design \citep{barreca2011saving}  for the gap it creates around the threshold in the distribution of the running variable.

Empirical researchers often use donut RD  designs when there are concerns that observations closest to the cutoff may be atypical in ways that threaten identification.\footnote{As of March 2026, a Google Scholar search for articles that contain the words ``donut'' and ``regression discontinuity'' returns almost 2,000 results.} A common motivation is the possibility of sorting or manipulation of the running variable, or administrative discretion in its measurement or recording, which can generate non-comparable units near the threshold. In other settings, heaping, rounding, or mass points in the running variable may induce irregular behavior near the cutoff that reflects reporting practices rather than the treatment assignment rule itself.

Depending on the context, researchers may view donut RD estimators either as more credible estimates of causal effects or as diagnostic tools. In the latter case, large differences between donut and conventional RD estimates are interpreted as evidence of violations of standard RD assumptions, without necessarily treating the donut estimate as a meaningful estimate of the causal parameter of interest. Conversely, small differences between donut and conventional RD estimates are often presented in robustness exercises as reassuring evidence that such identification concerns are limited.

Despite its popularity in empirical work, the donut RD approach is typically applied with little formal statistical guidance. In particular, it remains unclear how to interpret differences between conventional and donut RD estimates, or how large such differences must be to meaningfully support concerns about sorting, manipulation, or measurement problems at the cutoff. The consequences of trimming observations near the threshold for bias, variance, and confidence interval coverage are likewise poorly understood: excluding data close to the cutoff reduces local information and mechanically increases extrapolation, but the resulting trade-offs depend on trimming and bandwidth choices in ways that are rarely made explicit. Moreover, because conventional and donut estimators rely on largely overlapping samples and are therefore highly correlated, valid comparisons require careful treatment that is largely absent from applied practice.

To give a concrete example, Figure~\ref{fig:almond} shows average 1-year mortality rates for infants with birth weights around 1,500g, a “very low birth weight” threshold that often triggers additional medical interventions. A local linear RD regression,  as in \cite{almond2010estimating}, using a uniform kernel and a bandwidth of 85g yields a substantial and statistically significant RD estimate of 0.95\%, with a standard error of 0.22\%. \citet{barreca2011saving} argue that these results may be driven by heaping at 1,500g and 1,503g. These values correspond to 100g and one-ounce multiples, respectively, and observations rounded to these points may be systematically different from nearby observations.\footnote{For example, hospitals that round birthweights to the nearest 100g multiple, or have scales that measure birthweight in ounces rather than grams, may be systematically different from hospitals that report exact gram values.} Indeed, when a donut RD design excludes observations within 3g of the 1,500g threshold, the point estimate falls to 0.16\%, while the standard error increases to 0.28\%. Although this estimate appears qualitatively different from the conventional one, there is no formal framework for assessing whether the difference is sufficiently large to be attributed to random sampling variation alone. It is also unclear whether the donut RD estimator delivers a reliable estimate of a causal effect, or whether its value is driven by additional variance or extrapolation bias. Finally, it is not known whether the standard errors and confidence intervals reported by standard software packages for donut RD estimators are valid.

\begin{figure}[t]
	\centering
	\resizebox{.49\textwidth}{!}{\input{./Graphics/almond-normal.tex}}
	\resizebox{.49\textwidth}{!}{\input{./Graphics/almond-donut.tex}}
	\caption{Average 1-year mortality rates of infants with birth weights around 1,500g. Size of dots is proportional to number of observations and the total number of observations is 202,071. Left panel shows local linear RD fit with uniform kernel and bandwidth of 85g. Right panel shows fit of same specification when data points with birth weight between 1,497g and 1,503g are excluded from the data. \label{fig:almond}} %
\end{figure}

This paper provides econometric tools to address these questions. First, we show that donut RD estimators exhibit substantially higher bias and variance than conventional RD estimators when the standard RD assumptions hold. For example, under local linear estimation, excluding observations within 10\% of the main bandwidth around the treatment threshold increases bias by 41\% to 63\% and variance by 53\% to 61\%, depending on the kernel function. Second, we show that recently developed bias-aware confidence intervals \citep{armstrong2018optimal,armstrong2018simple, kolesar2018discrete} remain valid in donut RD designs without requiring special adjustments. However, trimming observations near the threshold can substantially increase the length of these confidence intervals: excluding observations within 10\% of the bandwidth increases asymptotic interval length by 26\% to 33\%, depending on the kernel. Third, we develop valid statistical tests for equality of conventional and donut RD estimands within a bias-aware framework that explicitly accounts for the dependence between the two estimators. Finally, we propose a new donut-based specification test with superior local power properties relative to existing approaches, at least against a class of plausible local alternatives.\footnote{The results developed in this paper extend directly to settings where the goal is to extrapolate treatment effects away from the cutoff.}

The remainder of the paper is organized as follows. Section~\ref{sec::setup} describes the general setup of conventional and donut RD designs. Sections~\ref{sec::point_estimation} and~\ref{sec::CIs} analyze point estimation and confidence intervals, respectively. Section~\ref{sec::specification_testing} studies donut RD methods as tools for specification testing. Section~\ref{sec::fuzzy} discusses extensions of our results to fuzzy RD designs. Section~\ref{sec::applications} presents two empirical illustrations, and Section~\ref{sec::conclusions} concludes.

\section{Setup}\label{sec::setup}

\subsection{Conventional RD Designs} Consider a conventional sharp regression discontinuity (RD) design in which the researcher is
interested in the causal effect of a binary treatment.\footnote{We focus on sharp RD designs in this section to simplify the exposition. See Section~\ref{sec::fuzzy} below for an extension of our results to fuzzy RD designs.} The data are
an independent sample $\{(Y_i,X_i,T_i), i \in [n]\}$ of size $n$ from some large population, where  $[n]= \{1, \dots, n\}$. Here
$Y_i\in\IR$ is the outcome of interest, $X_i\in\IR$ is the running variable, and
$T_i\in\{0,1\}$ is an indicator for the event that unit $i$ receives the treatment.
In a sharp RD design, units receive the treatment if and only if
the running variable exceeds some known threshold,
which we normalize to zero without loss of generality, so  $T_i=\1 {X_i \geq 0}.$
Writing $m_+ =\lim_{x\downarrow 0}m(x)$ and $ m_- =\lim_{x\uparrow 0}m(x)$
for the right and left limits, respectively, of a generic continuous function $m$ evaluated
at zero, we also denote the jump in the conditional expectation
of the observed outcome given the running variable at zero by
\begin{align*}
	\tau = \mu_{+} - \mu_{-}, \quad \mu(x) =\E(Y_i|X_i=x).
\end{align*}
In a potential outcomes
framework, the parameter $\tau$ coincides with the average treatment effect of units at the
cutoff under certain continuity conditions.
Local linear regression  \citep{fan1996local} has then arguably become the most commonly used estimation strategy  to estimate this parameter, due to its attractive theoretical properties and easy implementation. Specifically, the local linear estimator of $\tau$ is given by
$$\widehat\tau(h) = e_1^\top\argmin_{\beta}\sum_{i \in [n]} K_h(X_i)(Y_i- (T_i, X_i, T_iX_i, 1)^\top\beta)^2.$$
Here  $K$ is a kernel function with compact support, say $[-1,1]$, $h>0$ is a bandwidth, $K_h(x) = K(x/h)/h$, and $e_1 = (1,0,0,0)^\top$ is the first unit vector, whose role in the above formula is simply to extract the
appropriate coefficient from the right-hand side. Local linear regression thus proceeds by fitting linear specifications with different intercept and
slope on either side of the threshold by weighted least squares, giving non-zero weights to units
with running variable values $X_i\in[-h,h]$ only. We note that  by simple least squares algebra we can write the local linear RD estimator as weighted averages of the outcomes,
$$ \widehat\tau(h)=\sum_{i \in [n]} w_i(h)Y_i,$$
with weights that depend on the data through the realizations $\mathcal{X}_n=\{X_1,\ldots,X_n\}$ of the running variable only.

\subsection{Donut RD Estimator}

A  donut RD  estimator  is similar to a conventional one, except that it excludes data points immediately surrounding the treatment threshold. Here we focus  on the special case of a donut RD design in which the same bandwidth $h$ and donut size $d\in[0,h)$ are used on either side of the threshold.\footnote{We could easily accommodate asymmetric settings at the cost of a more involved notation. In particular, the donut could be one-sided, such that only observations above or below the threshold are excluded.}
The corresponding  local linear donut RD estimator is
$$\widehat\tau(h, d) = e_1^\top\argmin_{\beta}\sum_{i \in [n]}K(X_i/h)(Y_i-(T_i, X_i, T_iX_i, 1)^\top\beta)^2\1 {|X_i|\geq d}. $$
This corresponds again to fitting linear specifications via weighted least squares on either side of the threshold, but now non-zero weights are only given to units with $X_i$ taking values in the ``donut-shaped'' set $[-h,-d]\cup[d,h]$. The donut estimator nests the conventional one as a special case, i.e., $\widehat\tau(h, 0)=\widehat\tau(h)$, and it can also be written as weighted averages of the outcomes,
$$\widehat\tau(h, d)=\sum_{i \in [n]} w_i(h,d)Y_i,$$
with weights that satisfy $w_i(h,0)= w_i(h)$  and depend on the data through the realizations $\mathcal{X}_n=\{X_1,\ldots,X_n\}$ of the running variable only. Let $\mathcal X=\operatorname{supp}(X_i)$ denote the support of the
running variable.

\subsection{Donut RD Estimand}

If the conditions that allow for identification of causal effects in a conventional RD design were to hold, the donut RD estimator would merely be a very inefficient estimator of the parameter $\tau$, as the exclusion of data points around the cutoff mechanically increases both the bias (through additional extrapolation) and the variance (through the reduced number of data points) of the estimator. Donut RD approaches are therefore precisely  motivated by concerns that the observations closest to the cutoff may be atypical in a way that threatens  identification.

Depending on the empirical context, the use of donut RD methods typically takes one of two principal forms in empirical practice. Researchers might either use  $\widehat\tau(h, d)$ as a more credible estimator of a causal effect, and also  base inference on this estimator; or they might interpret the magnitude of the difference between $\widehat\tau(h, d)$ and $\widehat\tau(h)$ as a measure of the severity of the threats to the conventional RD design. Note that the latter perspective does not require one to take $\widehat\tau(h, d)$ as evidence about the causal parameter.

Both of these uses of donut RD designs are typically only motivated heuristically, and the formal statistical properties of the corresponding methods are largely unclear. To make progress, we consider a simple but general theoretical framework that is intended to be applicable in a wide range of empirical settings. Specifically, we  postulate that there exists a hypothetical  function $\mu^*(x)$ whose jump at zero is the target of the donut RD estimator. We denote this jump parameter by
$$\tau^* = \mu^*_+ - \mu^*_-.$$
We assume that the function $\mu^*(x)$ coincides with the conditional expectation function $\mu(x)=\E(Y_i|X_i=x)$ outside the donut, but not necessarily inside  it. That is, we assume that $\mu^*(x)=\mu(x)$ for $|x|\geq d$, but that this equality does not (necessarily) hold for $|x|<d$.
We also assume that the function $\mu^*(x)$ falls into the usual smoothness class of twice differentiable functions (except at the threshold) with bounded second derivatives, which is generally used to justify local linear regression approaches. That is, we assume that
\begin{align}
	\mu^*\in \mathcal{F}(M) = \{ m_1(x)\1{x\geq 0 }+ m_0(x)\1{x< 0 }, \|m_t''(\cdot)\|_{\infty} \leq M, t\in\{0,1\}\},
\end{align}
where $M>0$ is a uniform smoothness bound specified by the analyst. This setup implies that the function $\mu^*(x)$ is identified from the distribution of observable quantities for values of $x$ outside the ``donut hole'', and that the data are at least to some extent informative about the value of $\tau^*$ due to the shape restrictions imposed by the assumption that $\mu^*\in\mathcal{F}(M)$.
Depending on the empirical circumstances, the goal of a donut RD design could then be interpreted as  estimating $\tau^*$, or as formally testing the null hypothesis that $\mu(x)=\mu^*(x)$ for all $x$ by using a sample analogue $\tau^*-\tau$ as a test statistic.

\begin{example}[Causal Setup] Our  generic setup maps directly into a model where the validity of a causal design is undermined by the presence of some atypical observations near the cutoff only.
	Suppose there is a hypothetical sample $\{(Y_i(1),Y_i(0),X_i^*,X_i,T_i), i=1,\ldots,n\}$ of size $n$ from a large population, where  $Y_i(1)$ and $Y_i(0)$ are a unit's potential outcomes with and without receiving the treatment, respectively, so that $Y_i = Y_i(T_i)$; and  $X_i^*$ is  a ``natural'' running variable that would be observed in the absence of the possible data issues or the mechanism that induces sorting.
	The observed running  variable $X_i$ is further assumed to be identical to $X_i^*$ for those units whose realization of the latter falls outside the donut hole, and to fall into the donut if $X_i^*$ does so as well. That is,
	\begin{align*}
		X_i = X_i^* \textnormal{ if } |X_i^*|\geq d, \textnormal{ and } |X_i|<d \textnormal{ if } |X_i^*|<d,
	\end{align*}
	Treatment assignment is
	based on the observed running variable, so that  $T_i =\1{X_i\geq 0}$, and the observed outcome
	is $Y_i = Y_i(T_i)$. The parameter of interest is the average treatment effect among units
	whose ``natural'' value of the running variable is at the treatment threshold:
	$$\tau^* = \E(Y_i(1)-Y_i(0)|X_i^*=0) =\mu^*_+ - \mu^*_-,\quad \mu^*(x) = \E(Y_i|X_i^*=x),$$
	which is generally different from $\tau=\mu_+ - \mu_-$, the jump in the conditional expectation $\E(Y_i|X_i=x)$
	of the observed outcome given the observed running variable at zero.
\end{example}

\begin{example}[Diagnostic Setup]
	Our general framework can also be interpreted purely diagnostically, without attaching a
	causal meaning to the donut RD estimand $\tau^*$. Suppose that there exists a function
	$\mu^*(x)\in\mathcal{F}(M)$ such that $\mu^*(x)=\mu(x)$ for all $|x|\ge d$, but
	$\mu^*(x)$ may differ from $\mu(x)$ for $|x|<d$. Here $\mu^*$ is not assumed to describe
	the conditional expectation in the absence of any data problem. Rather, it is simply a
	smooth benchmark function that agrees with the observed conditional expectation outside
	the donut and extends the pattern observed there into the neighborhood of the cutoff.
	In this setup, $\tau^*=\mu_+^*-\mu_-^*$ is the jump implied by the benchmark function
	$\mu^*$, while $\tau=\mu_+-\mu_-$ is the jump in the observed conditional expectation.
	If the conventional RD specification is correct, the same local pattern governs the
	data both inside and outside the donut, i.e.\ if $\mu(x)=\mu^*(x)$ for all $|x|<d$. Under
	this null, $\tau=\tau^*$. Conversely, if $\tau\neq\tau^*$, then the behavior of the
	conditional expectation within the donut differs from the behavior outside the donut in a
	way that is not compatible with the maintained smoothness restriction. In that sense,
	$\tau^*-\tau$ can be viewed as a measure of local misspecification of the conventional RD
	design, even when $\tau^*$ is not itself an estimand of substantive interest.\footnote{
	For example, hospitals may round birthweights at focal values near 1{,}500g, so that
	mortality behaves differently for observations very close to the threshold than for
	observations slightly farther away. In that case, a discrepancy between $\tau$ and $\tau^*$
	indicates that the local pattern near the cutoff differs from the pattern outside the donut,
	without requiring that the donut RD estimand $\tau^*$ itself have a causal interpretation.}\end{example}

\subsection{Two Asymptotic Regimes}

For our large sample analysis of econometric methods in donut RD designs, we consider two different asymptotic regimes that present alternative perspectives on the size of the donut $d$ relative to the bandwidth $h$. We note that practitioners are not meant to explicitly choose between the two regimes, and that most methods we present in this paper work under either regime.

\begin{assumption}[Small Donut]\label{ass:sd} The donut size satisfies $d = c h $ for some $c\in[0,1)$,  $h\to0$ and $nh\to\infty$ as $n\to \infty$.
\end{assumption}

We refer to the framework introduced in Assumption~\ref{ass:sd} as ``small donut asymptotics''. It models the size of the donut as proportional to the bandwidth, which in turn tends to zero at an appropriate rate. Small donut asymptotics are meant to deliver accurate approximations in commonly found empirical settings that remove observations from an area around the cutoff that is small relative to the bandwidth. Note that while this framework formally allows for $c\in[0,1)$, the resulting asymptotic approximations only tend to be reliable for values of $c$ closer to zero, say $c\in[0,.2)$.

\begin{assumption}[Large Donut]\label{ass:ld}
	The donut size $d$ is fixed, $h\to d$ and $n(h-d)\to\infty$ as $n\to \infty$.
\end{assumption}

Assumption~\ref{ass:ld} is an alternative ``large donut'' asymptotic framework that treats $d$ as fixed and the bandwidth as approaching $d$ at an appropriate rate. This would be appropriate for settings in which the range of the support of the running variable that is used for estimation is much smaller than the gap created by the donut that needs to be extrapolated.\footnote{\cite{dowd2021donuts} considers a related setting, but focuses only on inference for the identified set rather than the jump at the cutoff, which is typically the main parameter of interest and the focus of our analysis.}

\subsection{Further Regularity Conditions} In addition to the two asymptotic frameworks above, we also introduce some further, and largely standard, regularity conditions.

\begin{assumption}\label{ass:rv} The running variable $X_i$ is continuously distributed with density function $f$ that is continuous on either side of the cutoff,  bounded and bounded away from zero over an open interval that contains the donut hole.
\end{assumption}

Continuity of the running variable is often assumed in the RD literature, although it is not necessary for valid estimation and inference based on local linear regression \citep{armstrong2018optimal,kolesar2018discrete}. We still maintain this assumption throughout the paper, as it often simplifies the derivations of analytic expressions for  asymptotic approximations.

\begin{assumption}\label{ass:moments}\begin{enumerate*}[label=(\roman*)]		\item For all $x\in \mathcal{X}$ and some $q>2$, $\E[|Y_i - \E[Y_i |X_i]|^q| X_i=x]$ exists and is uniformly bounded;
		\item $\V[Y_i |X_i=x]$ is $L$-Lipschitz continuous for all $x \in \mathcal X \setminus \{0\}$ and uniformly bounded away from zero.
	\end{enumerate*}
\end{assumption}

These conditions are needed in order to apply a central limit theorem in various places.\footnote{These conditions are stronger than necessary for our results on point estimation and confidence intervals below, and only need to hold for $x\notin(-d,d)$ in these cases. The assumption is required as stated for our results on specification testing.}

\begin{assumption}\label{ass:kernel}
	The kernel function $K$ is a bounded and symmetric density function with support contained in $[-1,1]$. In addition, $K$ is continuously differentiable on $(0,1)$.
\end{assumption}

This assumption is satisfied by most standard kernel functions, like the uniform, triangular, or Epanechnikov  kernel, for example. Kernel functions with unbounded support, like the Gaussian kernel, could be accommodated at the cost of additional algebra in some of the proofs. We also define the following functionals of the kernel:
\begin{align*}
	&B_K(c) = \int_{c}^1 J_K(u,c) K(u) u^2 du, \quad
	S_{K}(c)=\int_{c}^1 J_K(u,c)^2K(u)^2 du, \\ &J_K(u,c)  = e_1^\top\left(\int_c^1 (1,t)^\top (1,t) K(t) dt\right)^{-1}(1,u)^\top,
\end{align*}
for any constant $c \in [0,1)$.

\section{Point Estimation}\label{sec::point_estimation}

In this section, we study the properties of $\widehat\tau(h, d)$ as a point estimator of $\tau^*$.  Our main result shows that under small donut asymptotics the donut RD estimator stays consistent and asymptotically normal, but that the size of the donut can substantially affect the magnitudes of the asymptotic bias and variance. To state the result, we write the bias and variance of $\widehat\tau(h, d)$ conditional on the realizations $\mathcal{X}_n=\{X_1,\ldots,X_n\}$ of the running variable as  $b(h, d) = \E(\widehat\tau(h, d)|\mathcal{X}_n) -\tau^* $ and $s^2(h, d) = \V(\widehat\tau(h, d)|\mathcal{X}_n)$, respectively, and note that these terms can be written as
\begin{align*}
	b(h, d) = \sum_{i \in [n]}  w_{i}(h,d)\mu(X_i)-\tau^*\textnormal{ and }
	s^2(h,d) = \sum_{i \in [n]} w_i(h,d)^2 \sigma_{i}^2,
\end{align*}
with $\sigma_i^2 = \V(Y_i|X_i)$ the conditional variances of outcomes given their corresponding running variable values.
The following theorem gives asymptotic approximations to these quantities.

\begin{theorem}\label{theorem:pointest_smalldonut}
	Suppose that   Assumption~\ref{ass:sd}   and   Assumptions~\ref{ass:rv}--\ref{ass:kernel} hold. 
	Suppose further that, on each side of the cutoff, $\mu^*$ is twice continuously differentiable up to the cutoff, with uniformly bounded second derivative. Then
	\begin{gather*}\frac{\widehat\tau(h, d)-b(h, d)-\tau^*}{s(h, d)}\stackrel{d}{\to} N(0,1), \textnormal{ where }\\
		b(h, d) =  h^2 B_K(c)\frac{\mu_{+}^{*\;''}  - \mu_{-}^{*\;''}  }{2}  + o_P(h^2) \textnormal{ and }  s^2(h, d) = \frac{1}{n h}  S_{K}(c) \left(\frac{\sigma^2_{+}}{f_{+}} +\frac{\sigma^2_{-}}{f_{-}} \right) + o_P\left(\frac{1}{n h}\right).
	\end{gather*}
\end{theorem}

\begin{figure}
	\centering
	\resizebox{.49\textwidth}{!}{\includegraphics{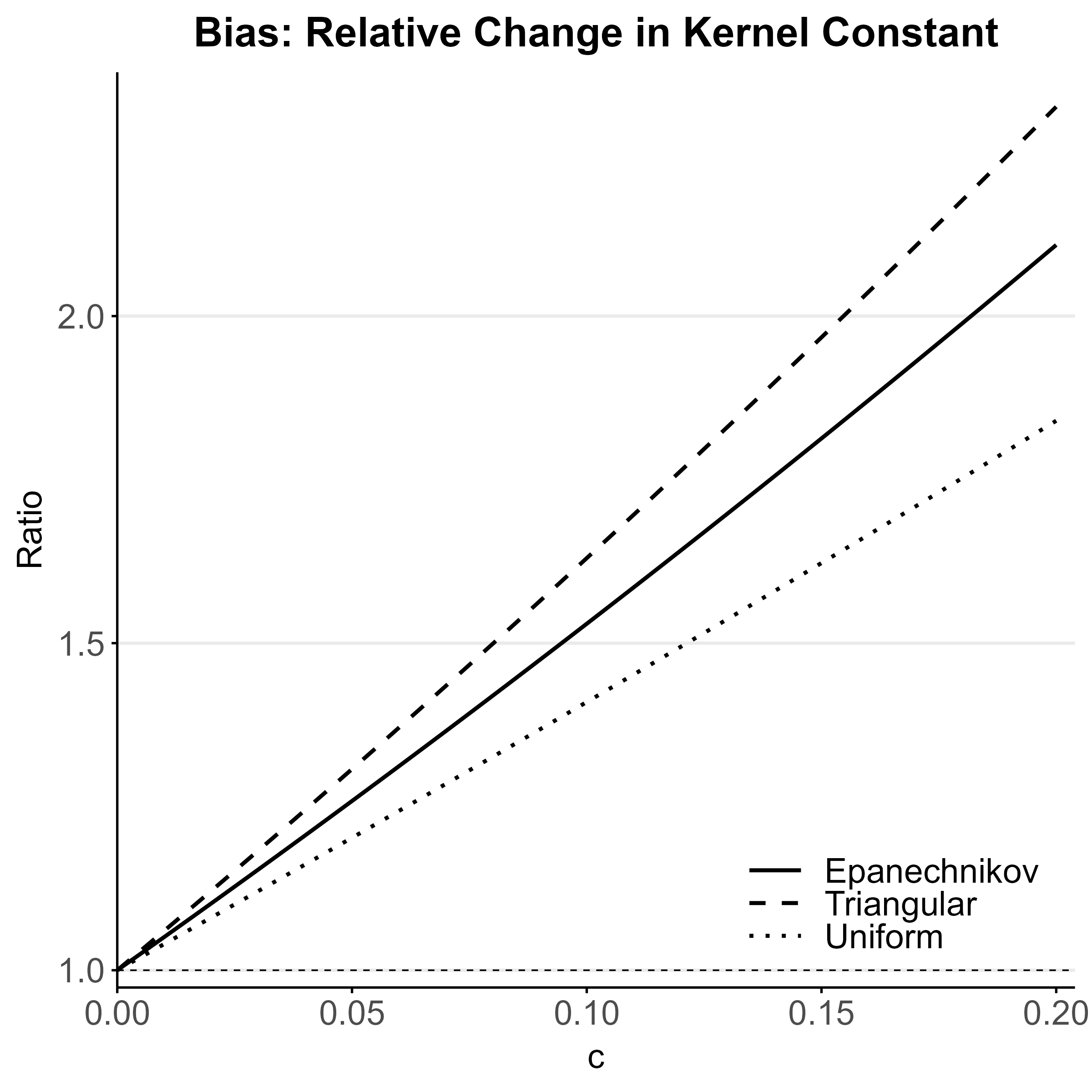}}
	\resizebox{.49\textwidth}{!}{\includegraphics{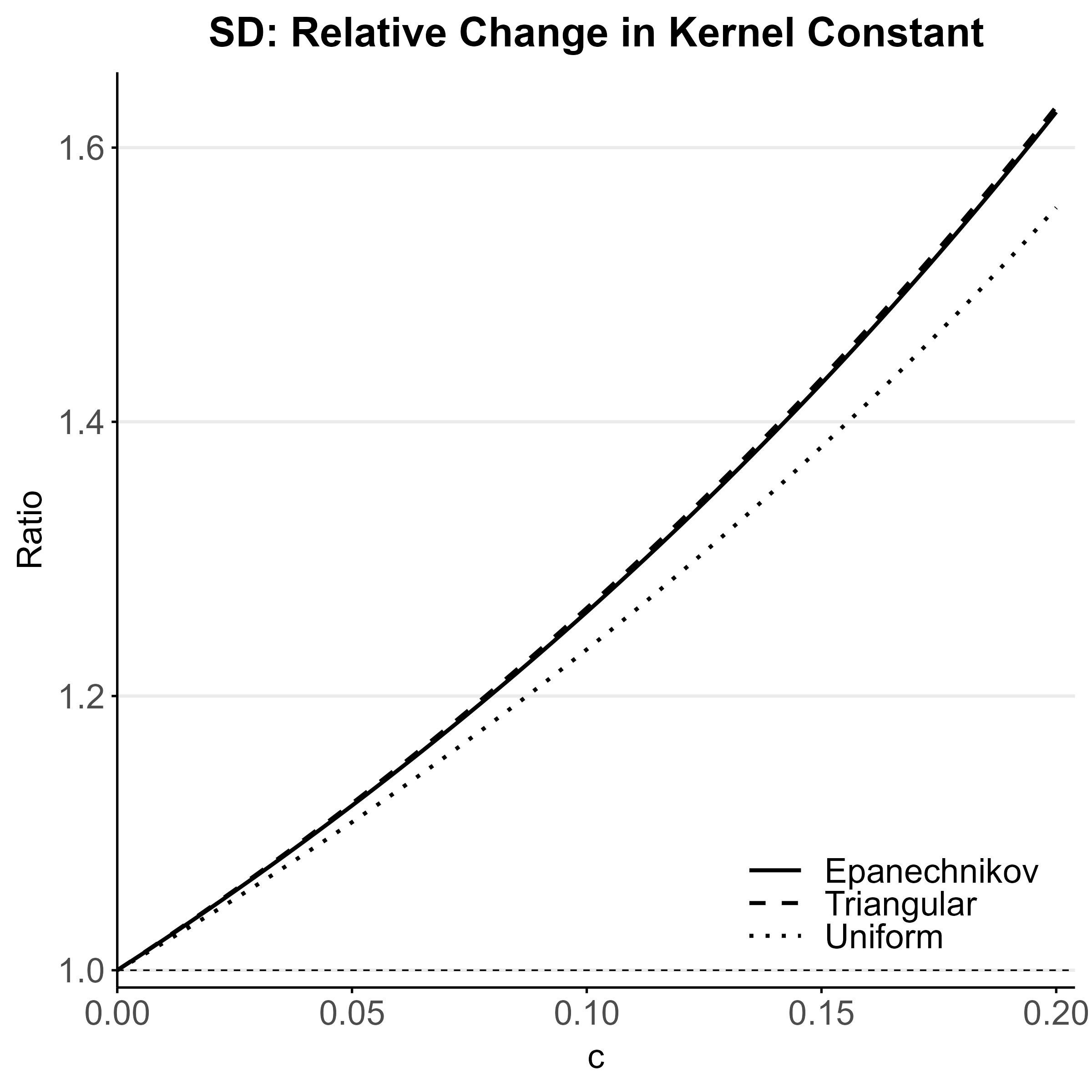}}
	\caption{Ratios of kernel constants in the asymptotic bias and standard deviations of the donut and the conventional RD estimator  for uniform, triangular, and Epanechnikov kernels.}\label{fig:bias_var_const}
\end{figure}

The result shows that the size of the donut affects the asymptotic bias and variance of the donut RD estimator through the kernel constants $B_K(c)$ and $S_{K}(c)$ only. The proof is straightforward and follows from observing that $\widehat\tau(h, d)$ is equal to a conventional RD estimator with a modified kernel function $K_c(u)=K(u)\1{|u|>c}$. The important insight  is that $B_K(c)$ and $S_{K}(c)$ can differ quite substantially from their ``no donut'' counterparts $B_K(0)$ and $S_K(0)$, even for moderate values of $c$. To illustrate this, Figure~\ref{fig:bias_var_const} plots the ratios $B_K(c)/B_K(0)$ and $\sqrt{S_{K}(c)/S_K(0)}$ for $c\in(0,.2)$ and three commonly used kernel functions. We can see, for example, that even with a moderate value such as $c=.1$, which corresponds to a donut RD that removes the observations that  differ by less than 10\% of the chosen bandwidth from the cutoff value, the absolute bias increases by 41\% to 63\%, the standard deviation increases by 
23\% to 26\%, and the variance increases by 53\% to 61\%, depending on the kernel function used.\footnote{
For illustration, we also plot the absolute values of $B_{K}(c)$ and  $S_{K}(c)$ in Figure~\ref{fig:full_bias_var_const} in the appendix.}

To give a point of reference for these numbers, note that reducing the bandwidth of the conventional RD estimator by 10\%, which removes observations within two slices of width $.1h$ on the outside rather than the inside of the estimation window, reduces the asymptotic bias by $1- (.9h)^2/h^2 =19\%$, and increases the variance by only $(n.9h)^{-1}/(nh)^{-1}-1\approx 11\%$.

\begin{remark}[Optimal Bandwidth]
	Given the characterization of the asymptotic bias and variance in Theorem~\ref{theorem:pointest_smalldonut}, the bandwidth that minimizes the asymptotic mean squared error (MSE) of the donut RD estimator for any particular donut size $d$ is given by:
	\[
	h_{\textnormal{MSE}}(d)
	= \argmin_{h > d} \left\{  h^4 B_K(d/h)^2 \left( \frac{\mu_{+}^{*\;''} - \mu_{-}^{*\;''}  }{2}\right)^2  +  \frac{1}{nh}  S_K(d/h) \left(\frac{\sigma^2_{+}}{f_{+}} +\frac{\sigma^2_{-}}{f_{-}} \right) \right\}.
	\]
	It is generally challenging to characterize how $h_{\textnormal{MSE}}(d)$ changes with $d$ as the  bandwidth enters the MSE formula not
	only as a multiplicative factor preceding the kernel constants, but also
	within the constants themselves.  Nonetheless,  one
	can show that $h_{\textnormal{MSE}}(d)$ is monotonically increasing in $d$ for a class of kernel functions that includes the three kernel
	functions considered above. This is not obvious a priori, as both the asymptotic bias and the asymptotic variance are increasing in $d$, and hence the direction of the change in mean squared error is not clear.

	\begin{proposition}\label{prop:bandwidth}
		Suppose that $d=d_0\times n^{-1/5}$ for some constant $d_0>0$,  that $\mu_+^{\star\; ''}\neq \mu_-^{\star\; ''}$, that Assumptions~\ref{ass:rv}--\ref{ass:kernel} hold, and that the function
		\[
		G_K(c) =
		\frac{S_K(c)+cS_K'(c)}
		{2B_K(c)\bigl(2B_K(c)-cB_K'(c)\bigr)}
		\]
		is strictly positive and strictly increasing on $(0,1)$, and satisfies
		\[
		\lim_{c\downarrow 0}c^5G_K(c)=0,
		\qquad
		\lim_{c\uparrow 1}c^5G_K(c)=\infty.
		\]
		Then $h_{\textnormal{MSE}}(d) = (d_0/c^*(d_0)) \times n^{-1/5}$,
		where
		\[
		c^*(d_0)=
		\argmin_{c\in(0,1)}
		\left(
		\left(\frac{d_0}{c}\right)^4
		B_K(c)^2
		\left(\frac{\mu_+^{\star\; ''}-\mu_-^{\star\; ''}}{2}\right)^2
		+
		\frac{c}{d_0}
		S_K(c)
		\left(
		\frac{\sigma_+^2}{f_+}+\frac{\sigma_-^2}{f_-}
		\right)
		\right)
		\]
		is uniquely defined. Moreover, the term $d_0/c^*(d_0)$ does not depend on $n$, is strictly increasing in $d_0$, and is such that
		\[
		\lim_{d_0\to 0}
		h_{MSE}(d)
		=
		\left(
		\frac{
		S_K(0)\left(\frac{\sigma_+^2}{f_+}+\frac{\sigma_-^2}{f_-}\right)
		}{
		B_K(0)^2(\mu_+^{\star\; ''}-\mu_-^{\star\; ''})^2
		}
		\right)^{1/5}
		\times n^{-1/5}.
		\]
	\end{proposition}

	Proposition~\ref{prop:bandwidth} establishes that the MSE-optimal bandwidth $h_{\textnormal{MSE}}(d)$ is increasing in the donut size under the high-level condition that the kernel-induced function $G_K(c)$ is strictly positive and strictly increasing on $(0,1)$, with the stated boundary behavior of $c^5G_K(c)$. For the uniform, triangular, and Epanechnikov kernels, this condition can be verified analytically by direct calculation (details are given after the proof of Proposition~\ref{prop:bandwidth} in the Appendix).  We conjecture that the same condition can also be checked, on a case-by-case basis, for other commonly used kernel functions.\footnote{If one implements a donut RD design by passing the subset of the data outside the donut to a conventional RD software package, the bandwidth selection algorithm might not automatically target the optimal bandwidth $h_{\textnormal{MSE}}(d)$ under the donut design if the algorithm relies on hard-coded values for the kernel constants.}
\end{remark}

\begin{remark}[Large Donut Asymptotics]\label{remark::largedonut} The case of large donut asymptotics is less interesting for point
	estimation as in this framework the parameter $\tau^*$ cannot be consistently estimated: even if we were to observe the full population distribution of $(Y_i,X_i)$ conditional on $|X_i|>d$, the shape restriction that $\mu^*\in\mathcal{F}(M)$ only yields that $\tau^*$ is partially identified and takes values in some identified set $T(M)$:
	$$\tau^*\in T(M) \equiv \left\{m_+ - m_-: m\in \mathcal{F}(M)\textnormal{ and } P( |m(X_i)- \mu(X_i)|\cdot\1{|X_i|\geq d}=0)=1 \right\}.$$
	With a continuously distributed running variable, the set $T(M)$ is an interval of length $2Md^2$ around the jump in the linear extrapolation of $\mu$ from the edge of the donut to the cutoff:
	$$T(M) = [\tau_\textnormal{Lin}( d) \pm M d^2],\qquad
	\tau_\textnormal{Lin}( d)= \mu(d)- \mu(-d) - d(\mu'(d) +\mu'(-d)).$$
	The estimator $\widehat\tau(h, d)$ simply  converges in probability to the midpoint of the identified set, i.e.\ $\widehat\tau(h, d)  =\tau_{\textnormal{Lin}}( d)+o_P(1)$, but that midpoint has no particular causal interpretation.
\end{remark}

\section{Confidence Intervals}\label{sec::CIs}

In this section, we consider confidence intervals for $\tau^*$ based on the donut RD estimator $\widehat\tau(h, d)$. In particular, we argue that  recently developed ``bias-aware'' confidence intervals \citep{armstrong2018optimal,armstrong2018simple, kolesar2018discrete} are asymptotically valid in donut RD designs under either large or small donut asymptotics without any special adjustment. The main insight is that the length of these confidence intervals can increase substantially with the size of the donut.

To describe ``bias-aware'' confidence intervals, note that it follows directly  from \citet{armstrong2018optimal} and \citet{kolesar2018discrete} that the conditional bias of the donut RD estimator is bounded  uniformly over $\mathcal{F}(M)$ by a term $\bar{b}(h, d) $ that can be explicitly calculated from the data:
$$\sup_{\mu^\star\in\mathcal{F}(M)}|b(h,d)|\leq  \bar{b}(h, d) \equiv  -\frac{M}{2}\sum_{i \in [n]} w_i(h,d)X_i^2\textnormal{sign}(X_i).$$
We can construct an estimator of the  conditional variance of the donut RD estimator as
$$\widehat s^2(h, d) =  \sum_{i \in [n]}  w_i(h,d)^2 \widehat\sigma_{i}^2,$$
where $\widehat\sigma_{i}^2$ is some suitable estimator of the conditional variance $\sigma_i^2 = \V(Y_i|X_i)$ such as the nearest-neighbor estimator studied by \citet{abadie2006large} and \citet{abadie2014inference}. We can then decompose the usual $t$-statistic as
\begin{align*}
	\frac{\widehat\tau(h,d)-\tau^*}{\widehat s(h,d)} = \frac{\widehat\tau(h,d)-\tau^* - b(h,d)}{\widehat s(h,d)} + \frac{b(h,d)}{\widehat s(h,d)}
\end{align*}
into the sum of a term that is approximately standard normal in large samples  and a term that can be bounded in absolute value by the observable quantity $\bar{b}(h,d)/\widehat s(h,d)$. This decomposition then leads to the bias-aware donut confidence interval with nominal confidence level $1-\alpha$,
$$C_n(d) =\left[\widehat\tau(h,d) \pm \textnormal{cv}_{1-\alpha}\left(\frac{\bar{b}(h,d)}{\widehat{s}(h,d)}\right)\widehat s(h, d) \right],$$
where  $\textnormal{cv}_{1-\alpha}(r)$ is the $1-\alpha$ quantile of  the $|N(r,1)|$ distribution, i.e., the distribution of the absolute value of a normal random variable with mean $r$ and variance 1. The next theorem shows that this result has correct asymptotic coverage uniformly over $\mu^*\in\mathcal{F}(M)$. 

\begin{theorem}\label{theorem:uniform_donut_estimator}
	Suppose that (i) either Assumption~\ref{ass:sd} or~\ref{ass:ld} holds; (ii) Assumptions~\ref{ass:rv}--\ref{ass:kernel} hold; and (iii) $\widehat s^2(h,d) =s^2(h,d)(1+o_{P}(1))$ uniformly over $\mathcal F(M)$. Then
	$$\liminf_{n\to\infty}\inf_{\mu^*\in\mathcal{F}(M)} P(\tau^*\in C_n(d)) \geq 1-\alpha. $$
\end{theorem}

\begin{figure}
	\centering
	\resizebox{.5\textwidth}{!}{\includegraphics{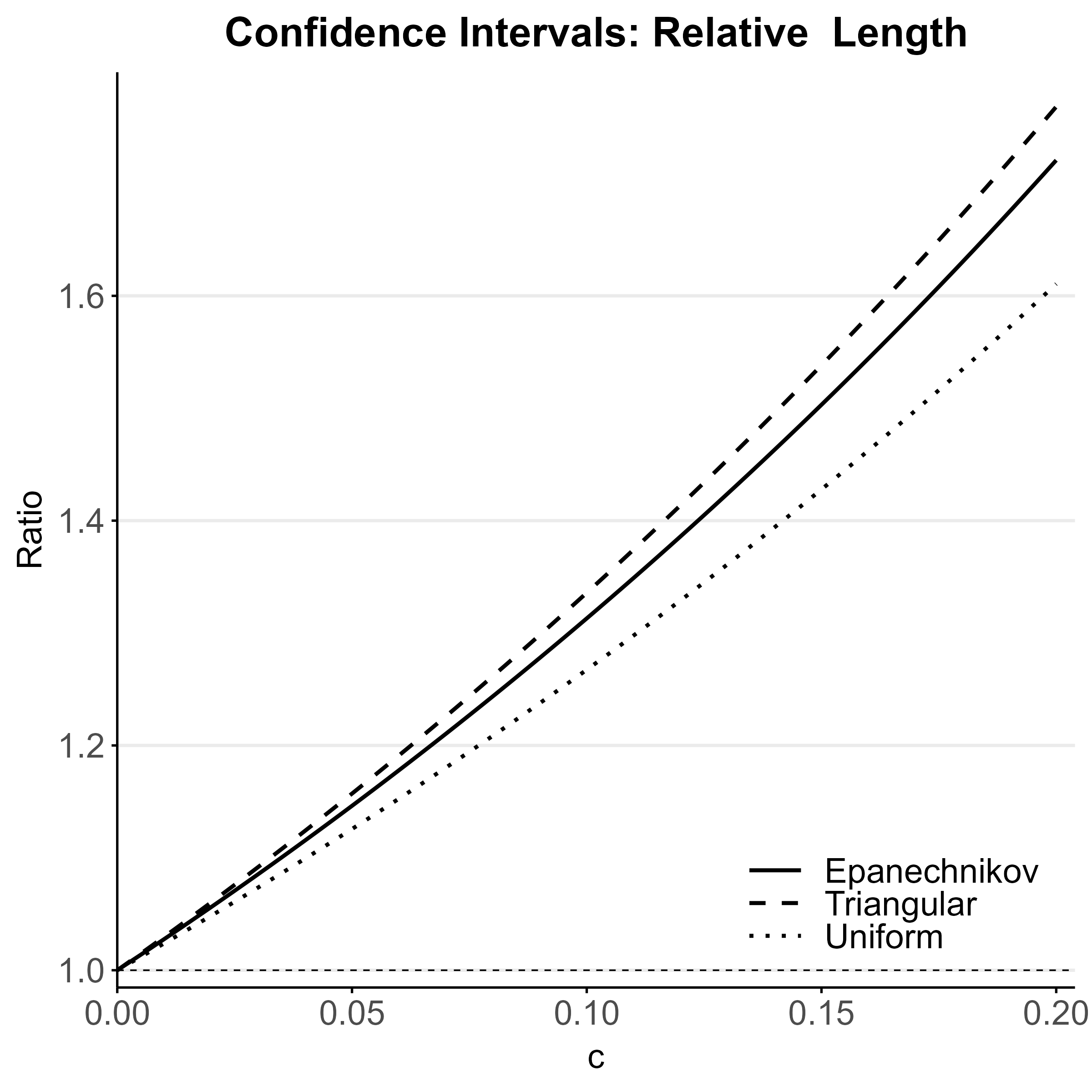}}
	\caption{Ratio of asymptotic length of ``donut'' and ``no donut'' confidence intervals as a function of  donut size for uniform, triangular, and Epanechnikov  kernel functions.\label{fig:asy.length.increase}}
\end{figure}

It is instructive to compare the length of the donut confidence interval $C_n(d)$ to that of its conventional counterpart $C_n\equiv C_n(0)$ in the case where $\mu$ equals $\mu^*$. Note that the presence of a donut  generally increases the standard error, which in turn widens the confidence interval, but it also affects the ``worst case bias to standard error'' ratio that appears inside the critical value function. To obtain an analytical result, it is convenient to choose the bandwidth such that it minimizes the ``worst case'' asymptotic MSE
of the conventional RD estimator, which means that one chooses the bandwidth
$$\bar{h}_{\textnormal{MSE}}=
\left(\frac{S_{K}(0) \left(\frac{\sigma^2_{+}}{f_{+}} +\frac{\sigma^2_{-}}{f_{-}} \right)}{4 B_K(0)^2  M^2}\right)^{\frac{1}{5}}\times n^{-1/5}.$$
This choice  implies in particular that $\bar{b}(\bar{h}_{\textnormal{MSE}},0)/\widehat{s}(\bar{h}_{\textnormal{MSE}},0)=1/2 + o_P(1)$. The presence of a donut then changes the length of the confidence interval by a factor of
\begin{align*}
	\frac{\textnormal{cv}_{1-\alpha}\left(\frac{\bar{b}(\bar{h}_{\textnormal{MSE}},d)}{\widehat{s}(\bar{h}_{\textnormal{MSE}},d)}\right)\widehat s(\bar{h}_{\textnormal{MSE}}, d)}{\textnormal{cv}_{1-\alpha}\left(\frac{\bar{b}(\bar{h}_{\textnormal{MSE}},0)}{\widehat{s}(\bar{h}_{\textnormal{MSE}},0)}\right)\widehat s(\bar{h}_{\textnormal{MSE}}, 0)}=\frac{\textnormal{cv}_{1-\alpha}\left(\frac{1}{2}\frac{B_K(c)}{\sqrt{S_K(c)}} \frac{\sqrt{S_K(0)}}{B_K(0)} \right) }{\textnormal{cv}_{1-\alpha}\left(\frac{1}{2}\right)}\cdot\sqrt{\frac{S_K(c)}{S_K(0)}} + o_P(1). \label{eq:asy.length.increase}
\end{align*}
We plot the relative increase in asymptotic length (that is, the first term on the right-hand side of the last equation) as a function of $c$ for different kernels in Figure~\ref{fig:asy.length.increase}. For example, with $c=.1$, which corresponds to a donut RD that removes the observations that differ by less than 10\% of the chosen bandwidth from the cutoff value, the length of the confidence interval increases by 26\% to 33\%, depending on the kernel function. The majority of this change can be attributed to the increase in the standard deviation, but the change in the critical value function adds to the length as well.

\begin{remark}[Robust Bias Correction] Confidence intervals based on ``robust bias correction'' \citep{calonico2014robust}  are a popular alternative to  ``bias-aware'' confidence intervals in conventional RD settings. In Appendix~\ref{appendix:addtheory}, we  show that RBC  confidence intervals only maintain correct coverage  under small donut asymptotics, but not under large donut asymptotics. We also show that their length is  generally even more affected by the size of the donut than that of bias-aware confidence intervals. This is because robust bias correction involves higher-order local polynomial estimation, and these higher-order local polynomials turn out to be more sensitive to donut trimming than local linear regression is.
\end{remark}

\section{Specification Testing}\label{sec::specification_testing}

Applied researchers often use donut RD designs as  diagnostic tools to investigate the validity of   conventional RD designs, taking ``large'' differences between the donut RD estimator and the conventional RD estimator as evidence of a violation of conditions that ensure identification of causal effects. This approach does not require that the donut RD estimator is by itself  a reasonable estimator of a causal effect.

Here we	consider two hypothesis tests (one formalizes a heuristic strategy commonly used in practice and one is based on a new approach) and compare their power properties over a class of reasonable local alternatives. We formalize the goal of testing for correct specification as the null hypothesis:
$$H_0: \mu^*(x) = \mu(x) \textnormal{ for all } |x|<d,$$
under our maintained conditions that  $\mu^*(x) = \mu(x)$ for all $|x|\geq d$, and $\mu^*\in\mathcal{F}(M)$.\footnote{Note that it does not suffice to state the null hypothesis as $H_0:\tau=\tau^*$, as some control of $\mu$ within the donut is necessary to derive the properties of the conventional RD estimator.} This formulation captures the idea that under the null hypothesis the observed data within the donut should be compatible with (i) the data outside the donut  and (ii) the shape constraints imposed by the assumption that $\mu^*\in\mathcal{F}(M)$. Throughout this section, we consider a nontrivial donut, so that $d>0$; under small-donut asymptotics, this additionally requires $c=d/h>0$.

\subsection{Conventional Strategy: Donut vs.\ Standard RD Estimates}

We first formalize a testing strategy based on comparing donut RD estimates to conventional RD estimates. We denote the difference between the two estimators by
\begin{align*}
	&\widehat\Delta(h, d) \equiv \widehat\tau(h, d)-\widehat\tau(h, 0)  = \sum_{i \in [n]} (w_i(h, d)-w_i(h, 0))Y_i,
\end{align*}
and the respective conditional bias and variance by $b_\Delta(h, d) = \E(\widehat\Delta(h, d)|\mathcal{X}_n)$ and $s_\Delta^2(h, d) = \V(\widehat\Delta(h, d)|\mathcal{X}_n)$, respectively. These can be written as
\begin{align*}
	b_\Delta(h, d) &= \sum_{i \in [n]}  (w_{i}(h,d)-w_{i}(h,0)) \mu(X_i)\textnormal{ and}\\
	s_\Delta^2(h,d) &= \sum_{i \in [n]}  \left( w_i(h,d)^2 + w_i(h,0)^2 -2 w_i(h,d) w_i(h,0)  \right) \sigma_{i}^2,
\end{align*}
respectively. We can then prove  the new result that under our null hypothesis the conditional bias term is bounded in finite samples uniformly over $\mathcal{F}(M)$ as follows
\begin{align*}
	\sup_{\mu^\star\in\mathcal{F}(M)} |b_\Delta(h, d)|\leq \bar{b}_\Delta(h, d)\equiv -\frac{M}{2}\sum_{i \in [n]} (w_i(h,d)-w_i(h,0))X_i^2 \sgn(X_i).
\end{align*}
On the other hand, a natural estimate of the conditional variance is given by:
$$\widehat s_\Delta^2(h,d) = \sum_{i \in [n]}  \left( w_i(h,d)^2 + w_i(h,0)^2 -2 w_i(h,d) w_i(h,0)  \right) \widehat\sigma_{i}^2,$$
with $\widehat\sigma_{i}^2$ again a nearest-neighbor estimate of $\sigma_i^2 =\V(Y_i|X_i)$. Following arguments analogous to those used to construct ``bias-aware'' confidence intervals, we use the decision rule to
\begin{align*}
	\textnormal{reject }H_0 \textnormal{ if } |t_\Delta| >\textnormal{cv}_{1-\alpha}\left(\frac{\bar{b}_\Delta(h, d)}{\widehat s_\Delta(h,d)} \right), \textnormal{ where }t_\Delta =\frac{\widehat\Delta(h,d)}{\widehat s_\Delta(h,d)}
\end{align*}
is a $t$-statistic that
is approximately normal in large samples with unit variance and mean bounded in absolute value by $\bar{b}_\Delta(h, d)/\widehat s_\Delta(h,d)$ under the null hypothesis.
The following theorem shows that the resulting test has correct size.
\begin{theorem}\label{theorem:standard_donut_test}
	Suppose that (i) either Assumption~\ref{ass:sd} or~\ref{ass:ld} holds; (ii) Assumptions~\ref{ass:rv}--\ref{ass:kernel} hold; and (iii) $\widehat s_\Delta^2(h,d) =s_\Delta^2(h,d)(1+o_P(1))$ uniformly over $\mathcal F(M)$. Then, under $H_0$:
	$$\limsup_{n\to\infty}\sup_{\mu^*\in\mathcal{F}(M)} P\left(|t_\Delta| \geq \textnormal{cv}_{1-\alpha}\left(\frac{\bar{b}_\Delta(h, d)}{\widehat s_\Delta(h,d)} \right)\right) \leq \alpha. $$
\end{theorem}

\begin{remark}[Variance Structure]The test statistic considered here is based on the difference of two generally highly correlated
	quantities. In particular, under small donut asymptotics the conditional variance  of $\widehat\Delta(h,d)$  satisfies:
	$$s_\Delta^2(h,d) = \frac{1}{nh} (S_{K}(c)+ S_K(0) -2\widetilde S_{K}(c))\left(\frac{\sigma^2_+}{f_+}+\frac{\sigma^2_-}{f_-}\right) +o_P\left(\frac{1}{n h}\right)$$
	where $S_{K}(c)$ is as defined above and
	$$\widetilde S_{K}(c) = \int_c^1 J_K(u,c)J_K(u,0)K(u)^2du$$
	captures the dependence between the conventional and the donut RD estimator.
\end{remark}

\subsection{Alternative Strategy: Donut vs.\ ``Within Donut'' RD Estimates}
The test statistic in the previous subsection is the difference of two potentially highly correlated quantities. An obvious alternative for testing $H_0$
is to compare the donut RD estimate to a conventional RD estimator with bandwidth $d$. We refer to this latter estimator $\widehat\tau(d, 0)=\widehat\tau(d)$ as the ``within donut'' RD estimator, as it only uses data within the donut hole. That is, one can base a  test of $H_0$ on (an appropriately studentized version of) the difference
\begin{align*}
	&\widehat\Gamma(h, d) \equiv \widehat\tau(h, d)-\widehat\tau(d, 0)  = \sum_{i \in [n]} (w_i(h, d)-w_i(d, 0))Y_i.
\end{align*}
This approach compares two statistically independent quantities, as these are based on non-overlapping subsets of the data. The  downside of this construction is that a large overall sample size is required to use this approach as the variance of $\widehat\tau(d, 0)$ is large and its distribution may not be well-approximated by a central limit theorem without
a sufficient number of data points within the donut.

We denote the respective conditional bias and variance by $b_\Gamma(h, d) = \E(\widehat\Gamma(h, d)|\mathcal{X}_n)$ and $s_\Gamma^2(h, d) = \V(\widehat\Gamma(h, d)|\mathcal{X}_n)$, which can be written as
\begin{align*}
	b_\Gamma(h, d) &= \sum_{i \in [n]}  (w_{i}(h,d)-w_{i}(d,0)) \mu(X_i)\textnormal{ and}\\
	s_\Gamma^2(h,d) &= \sum_{i \in [n]}  \left( w_i(h,d)^2 + w_i(d,0)^2  \right) \sigma_{i}^2,
\end{align*}
We then show that we can bound the bias in finite samples uniformly over $\mathcal{F}(M)$ under the null hypothesis by
\begin{align*}
	\sup_{\mu^*\in\mathcal{F}(M)} |b_\Gamma(h, d)| \leq \bar{b}_\Gamma(h, d) \equiv -\frac{M}{2}\sum_{i \in [n]} (w_i(h,d)-w_i(d,0))X_i^2 \sgn(X_i).
\end{align*}
On the other hand, a natural estimate of the conditional variance is  given by
$$\widehat s_\Gamma^2(h,d) = \sum_{i \in [n]}  \left( w_i(h,d)^2 + w_i(d,0)^2 \right) \widehat\sigma_{i}^2,$$
with $\widehat\sigma_{i}^2$ again a nearest-neighbor estimate of $\sigma_i^2 =\V(Y_i|X_i)$. Following arguments analogous to those used to construct ``bias-aware'' confidence intervals,  we use  the decision rule to
\begin{align*}
	\textnormal{reject }H_0 \textnormal{ if } |t_\Gamma| >\textnormal{cv}_{1-\alpha}\left(\frac{\bar{b}_\Gamma(h, d)}{\widehat s_\Gamma(h,d)} \right), \textnormal{ where }t_\Gamma =\frac{\widehat\Gamma(h,d)}{\widehat s_\Gamma(h,d)}
\end{align*}
is a $t$-statistic that is approximately normal in large samples with unit variance and mean bounded in absolute value by $\bar{b}_\Gamma(h, d)/\widehat s_\Gamma(h,d)$ under the null hypothesis.
The following theorem shows that the resulting test has correct size.

\begin{theorem}\label{theorem:alternative_donut_test}
	Suppose that (i) either Assumption~\ref{ass:sd} or~\ref{ass:ld} holds; (ii) Assumptions~\ref{ass:rv}--\ref{ass:kernel} hold; and (iii) $\widehat s_\Gamma^2(h,d) =s_\Gamma^2(h,d)(1+o_P(1))$ uniformly over $\mathcal F(M)$. Then, under $H_0$:	$$\limsup_{n\to\infty}\sup_{\mu^*\in\mathcal{F}(M)} P\left(|t_\Gamma| \geq \textnormal{cv}_{1-\alpha}\left(\frac{\bar{b}_\Gamma(h, d)}{\widehat s_\Gamma(h,d)} \right)\right) \leq \alpha. $$
\end{theorem}

\subsection{Power Comparisons}
We compare the power of the two tests in the previous subsections under small donut asymptotics\footnote{The case of small donut asymptotics is the more interesting and illustrative one as under large donut asymptotics the parameter $\tau^*$ is only partially identified, and hence no hypothesis test has non-trivial asymptotic power against any alternative inside the identified set.} with $h= h_0\times n^{-1/5}$ for some constant $h_0>0$ under sequences of local alternatives in which $\mu_n^*(x)$ differs from a continuous piecewise  quadratic baseline function $\mu(x)$ inside the donut by a piecewise quadratic perturbation that is continuous at
$\pm d$ and  discontinuous at the cutoff, and  chosen such that
$$\tau^*_n-\tau = \lambda /\sqrt{nh}$$
for some local parameter $\lambda\in\IR$.
That is, local alternatives are chosen such that the jump at the cutoff, which is the parameter that is targeted by donut RD estimation, scales with the asymptotic standard deviation of the test statistics. The local alternatives considered are
specifically of the form
$$H_{1,n}: \mu_n^*(x)=\mu(x)+ \eta_n(x) \textnormal{ for all } x \in \mathcal X,$$
where $\mu(x) = - M x^2\sgn(x)/2$ is chosen such that the conventional RD estimator achieves its ``worst case'' bias, and $ \eta_n(x)=\sgn(x)(\lambda /(2d^{2}\sqrt{nh}))\,(x-\sgn(x)d)^2\,\1{|x|<d}$.

\begin{figure}
	\centering
	\includegraphics[width=0.48\linewidth]{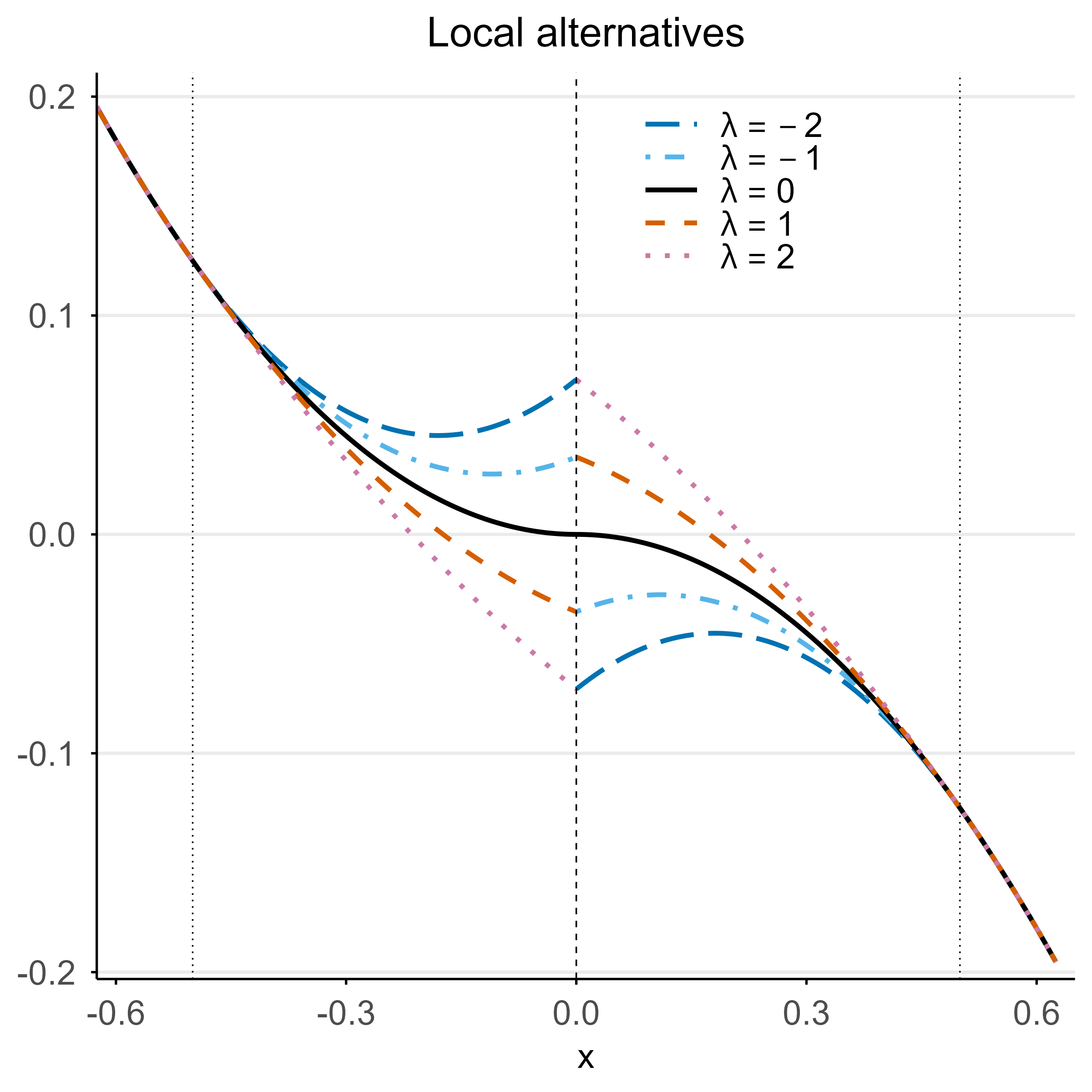}
	\caption{Illustration of local alternatives $\mu^\star_n(x)$ for different values of $\lambda$.}
	\label{fig:local_alternatives}
\end{figure}

\begin{theorem}\label{theorme::local_alternatives}
	Suppose that  (i) Assumption~\ref{ass:sd} holds and $c>0$; (ii) Assumptions~\ref{ass:rv}--\ref{ass:kernel} hold; (iii) $\widehat s_{\Delta}^2(h,d) =s_{\Delta}^2(h,d)(1+o_P(1))$ and  $\widehat s_{\Gamma}^2(h,d) =s_{\Gamma}^2(h,d)(1+o_P(1))$ uniformly over $\mathcal F(M)$; and (iv) the bandwidth is $h = h_0\times n^{-1/5}$. Then, under the local alternative $H_{1,n}$,
	$$ t_\Delta \overset{d}{\to} N(\mathcal D_\Delta(c,h_0,K) - \lambda\,\mathcal L_\Delta(c,K),1) \quad \text{ and } \quad  t_\Gamma\overset{d}{\to} N (\mathcal D_\Gamma(c,h_0,K) - \lambda\,\mathcal L_\Gamma(c,K),1)$$
	for constants
	\begin{gather*}
		\mathcal D_\Delta(c,h_0,K)
		=
		\frac{-M h_0^{5/2}\big(B_K(c)-B_K(0)\big)}
		{\sigma_\tau\sqrt{S_{K, \Delta}(c)}},
		\quad
		\mathcal D_\Gamma(c,h_0,K)
		=
		\frac{-M h_0^{5/2}\big(B_K(c)-c^2B_K(0)\big)}
		{\sigma_\tau\sqrt{S_{K,\Gamma}(c)}},\\
		\mathcal L_\Delta(c,K)
		=
		\frac{
		\int_0^c J_K(u,0)(u-c)^2K(u)\,du
		}{
		c^2 \sigma_\tau\sqrt{S_{K,\Delta}(c)}
		},
		\quad
		\mathcal L_\Gamma(c,K)
		=
		\frac{
		B_K(0) + 1 
		}{
		\sigma_\tau\sqrt{S_{K,\Gamma}(c)}
		},
	\end{gather*}
	with $\sigma_\tau^2=\sigma_+^2/f_+ + \sigma_-^2/f_-$ and $S_{K,\Delta}(c)=S_K(c)+S_K(0)-2\widetilde S_K(c)$ and $S_{K,\Gamma}(c)=S_K(c)+S_K(0)/c$.
\end{theorem}

Note that the constants $\mathcal D_\Delta(c,h_0,K)$ and $\mathcal D_\Gamma(c,h_0,K)$ capture the drift induced by the curvature of the baseline function $\mu(\cdot)$. They are the probability limits of the respective ``worst case bias to standard error'' ratios, and would also be present under the null hypothesis (which corresponds to $\lambda=0$). The constants $\mathcal L_\Delta(c,K)$ and $\mathcal L_\Gamma(c,K)$, on the other hand, capture the drift induced by the local alternative inside the donut.

\begin{figure}
	\centering
	\resizebox{.6\textwidth}{!}{		\includegraphics{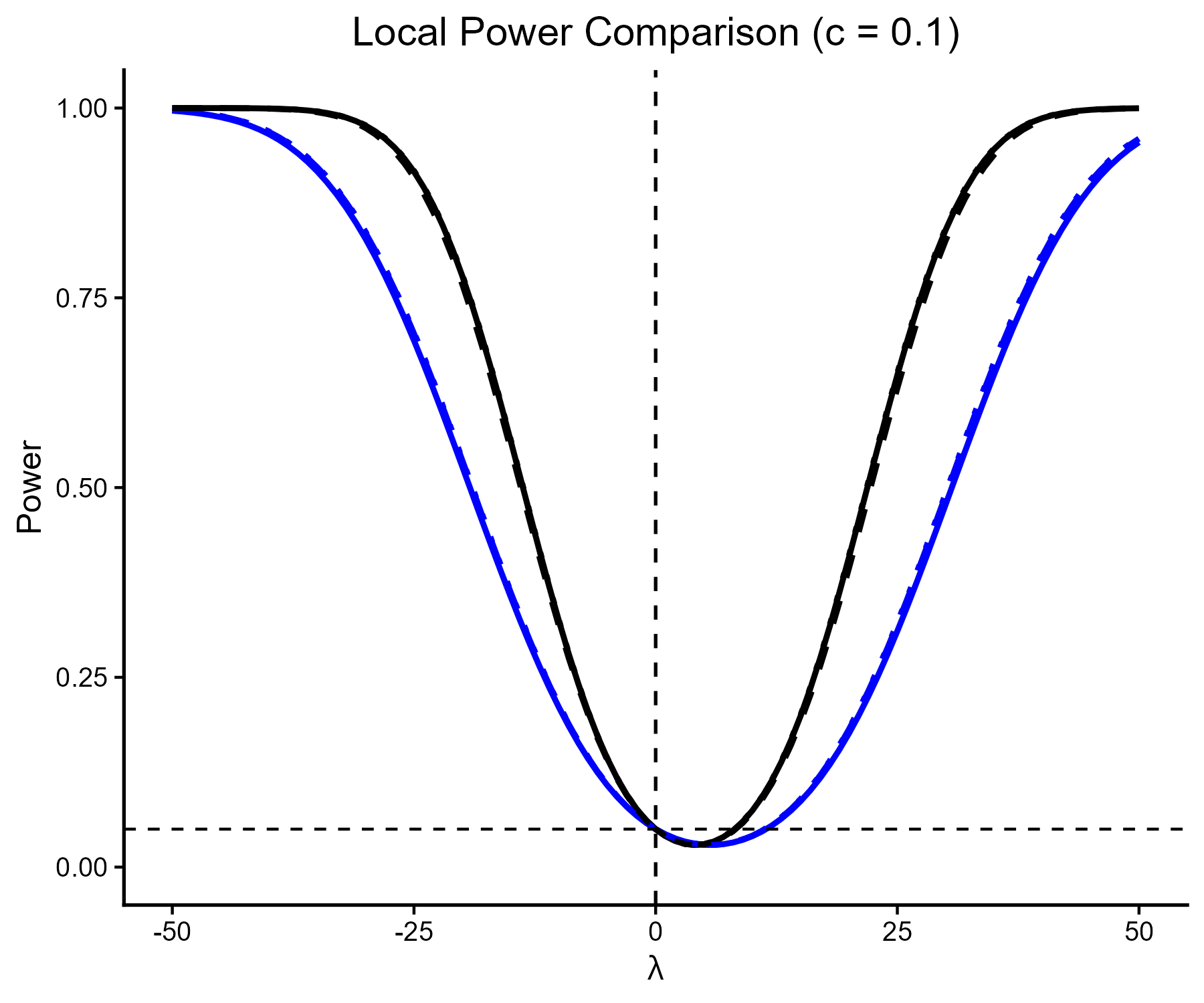}}
	\caption{Asymptotic power of the test $\Delta$ (blue) and $\Gamma$ (black) based on  local alternatives indexed by $\lambda$  for uniform ($\cdots$), triangular ($---$), and Epanechnikov ($-$) kernels and $\alpha=0.05$.}\label{fig:power}
\end{figure}

Theorem~\ref{theorme::local_alternatives} can be used to characterize the asymptotic power functions of the two tests under the local alternative $H_{1,n}$. It is convenient to choose  $h$ such that it minimizes the  ``worst case'' asymptotic MSE of the numerator of the respective $t$-statistic, as in this case the ``worst case bias to standard error'' ratios become asymptotically equal to $1/2$. With this choice of bandwidth, we have that
\begin{align*}
	\mathbb P\left(|t_\Delta| > \textnormal{cv}_{1-\alpha}\left(\frac{ \bar b_{\Delta}(h,d)}{s_{\Delta}(h,d)}\right) \Big|H_{1,n}\right)& = \mathbb P\left(\left|Z-\lambda\mathcal L_\Delta(c,K)\right| > \textnormal{cv}_{1-\alpha}\left(\frac{1}{2} \right)\right)+ o(1), \\
	\mathbb P\left(|t_\Gamma| > \textnormal{cv}_{1-\alpha}\left(\frac{ \bar b_{\Gamma}(h,d)}{s_{\Gamma}(h,d)}\right) \Big|H_{1,n}\right)& = \mathbb P\left(\left|Z-\lambda\mathcal L_\Gamma(c,K)\right| > \textnormal{cv}_{1-\alpha}\left(\frac{1}{2} \right)\right)+ o(1),
\end{align*}
where $Z\sim N(1/2,1)$ is a generic normal random variable with mean 1/2 and variance 1.
The first terms on the right-hand side of the last two displays are the asymptotic power functions of the two tests we consider. It is tedious to further characterize these functions analytically. We show their numerical values as a function of $\lambda$ for various kernels and the special case $c=.1$ in
Figure~\ref{fig:power} (the overall shape of the power curves is qualitatively very similar for other values of $c$). This plot clearly shows that the test $\Gamma$, which is based on our new strategy to compare donut and ``within donut'' estimates, generally dominates the test~$\Delta$, which formalizes the comparison of donut and conventional estimates that is commonly carried out in the empirical literature. This dominance result is of course tied to the specific form of the local alternative.

\section{Fuzzy Regression Discontinuity Designs}\label{sec::fuzzy}

In this section, we briefly outline how our analysis can be extended to fuzzy regression discontinuity (RD) designs; additional details are provided in Appendix~\ref{appendix:fuzzy}. In a fuzzy RD design, treatment assignment is determined by whether the running variable exceeds a cutoff, but compliance with the assignment rule is imperfect. As a result, the probability of treatment exhibits a discontinuity at the cutoff, but it might not jump from zero to one.

The parameter of interest is typically defined as the ratio of two sharp RD parameters,
\[
\theta = \frac{\tau_Y}{\tau_T}, \qquad
\tau_Y = \mu_{Y+} - \mu_{Y-}, \quad \tau_T = \mu_{T+} - \mu_{T-},
\]
provided $\tau_T\neq0$. Here, for a generic random variable $W_i$, we write $\mu_W(x) = \mathbb{E}[W_i \mid X_i = x]$ for its conditional expectation given the running variable.\footnote{Throughout this section, notation parallels that used earlier, with subscripts $Y$ and $T$ denoting outcomes and treatment indicators, respectively.} Under standard continuity assumptions \citep{hahn2001identification, dong2014alternative}, $\theta$ corresponds to the average causal effect at the cutoff for units whose treatment status is affected by the assignment rule (i.e., compliers).

A fuzzy donut RD estimator can be constructed as the ratio of two sharp donut RD estimators,
\[
\widehat{\theta}(h,d) = \frac{\widehat{\tau}_Y(h,d)}{\widehat{\tau}_T(h,d)}, \qquad
\widehat{\tau}_Y(h,d) = \sum_{i \in [n]} w_i(h,d) Y_i, \quad
\widehat{\tau}_T(h,d) = \sum_{i \in [n]} w_i(h,d) T_i,
\]
if $\widehat{\tau}_T(h,d) \neq 0$.
We then postulate that there are functions $\mu_Y^*\in\mathcal{F}(M_Y)$ and $\mu_T^*\in\mathcal{F}(M_T)$ that coincide with $\mu_Y$ and $\mu_T$ outside the donut, but possibly not inside it. The numerator and denominator of the fuzzy RD estimator are then interpreted as targeting the jumps $\tau_Y^*=\mu^*_{Y+} - \mu^*_{Y-}$ and $\tau_T^*=\mu^*_{T+} - \mu^*_{T-}$, respectively; and the fuzzy donut RD estimator itself is interpreted as targeting their ratio $\theta^* = \tau_Y^*/\tau_T^*$ provided that $\tau_T^* \neq 0$.

If the discontinuity in treatment probabilities is bounded away from zero outside the donut, it follows from a delta method argument that under small-donut asymptotics the fuzzy donut RD estimator is asymptotically equivalent to a sharp donut RD estimator with an appropriately defined outcome variable:
\[
\widehat{\theta}(h,d) - \theta \approx \sum_{i \in [n]} w_i(h,d) U_i,
\quad
U_i = \frac{Y_i - \tau_Y}{\tau_T} - \frac{\tau_Y}{\tau_T^2}\bigl(T_i - \tau_T\bigr).
\]
Under these conditions, all of our results for point estimation, confidence intervals, and specification testing under small-donut asymptotics apply directly to $\widehat{\theta}(h,d)$ after replacing the outcome variable $Y_i$ with the auxiliary outcome $U_i$. The corresponding methods can be made feasible by replacing the unobserved $U_i$ with plug-in estimates of the form $\wh U_i=(Y_i-\wh\theta(h,d) T_i)/\wh\tau_T(h,d)$.

This approach does not extend to large donut asymptotics though, as in this case both $\widehat{\tau}_Y(h,d)$ and $\widehat{\tau}_T(h,d)$, and consequently also $\widehat{\theta}(h,d)$, are generally inconsistent for their respective target parameters, and hence the delta method argument breaks down. For point estimation, a partial identification argument analogous to that in Remark~2 applies in this case. 
To construct valid confidence sets for the corresponding fuzzy RD estimand $\theta^\ast$, we recommend using the Anderson--Rubin--type procedure proposed by \cite{noack2024bias}. This approach is valid under both small- and large-donut asymptotics and does not require the treatment probability discontinuity to be bounded away from zero. For specification testing in fuzzy RD designs, we recommend conducting separate sharp RD specification tests for the numerator and denominator of the fuzzy RD estimator. The null hypothesis of correct specification is then rejected if the smaller of the two resulting $p$-values falls below the Bonferroni-adjusted significance level $\alpha/2$. Again, we provide details in  Appendix~\ref{appendix:fuzzy}.

\section{Two Empirical Applications}\label{sec::applications}

In this section, we present two empirical applications that illustrate our methods.

\subsection{Teacher contracts} We first revisit the data from \citet{bau2020teacher}, who study the effect of teachers' contract type, which can be either temporary or permanent, on various outcomes such as teachers' wages and teacher value added (TVA) in Pakistan. They exploit a policy shift in 2002 which led to teachers being hired almost exclusively on non-tenured temporary contracts. This policy shift was immediately preceded, however, by a budget crisis that resulted in a uniquely long period of low hiring from 1998 to 2001. \cite{bau2020teacher} conduct a (one-sided) fuzzy donut RD analysis with data from eight years before and four years after the reform while excluding the four-year period of low hiring.
\begin{figure}[h]
	\centering
	\resizebox{.49\textwidth}{!}{\input{./Graphics/bau_normal.tex}}
	\resizebox{.49\textwidth}{!}{\input{./Graphics/bau_donut_y.tex}}
	\resizebox{.49\textwidth}{!}{\input{./Graphics/bau_normal_treatment.tex}}
	\resizebox{.49\textwidth}{!}{\input{./Graphics/bau_donut_d_treatment.tex}}
	\caption{Average log salaries and fraction of temporary contracts around the discontinuity in the year of the reform. Size of dots is proportional to the number of observations and the total number of observations is 1,069. Left panel shows local linear RD fits with a uniform kernel using all available observations. Right panel shows the same specification when data points within four years below the cutoff are excluded (“donut hole”). 
	}\label{fig:bau}
\end{figure}

For illustration, we focus here on teachers' log monthly wages as the outcome.
Figure~\ref{fig:bau} shows the data together with the first-stage and reduced-form local linear regression fits in the left panel and the corresponding donut fits in the right panel.
 We can see that donut trimming has a particularly substantial effect on the first stage estimate of the jump in treatment probabilities.

As the donut size here is  exactly half of the effective bandwidth, we consider this to be a  setting where small donut asymptotics are less plausible and large donut asymptotics deliver more reasonable approximations. In Table~\ref{tab:bau}, we report fuzzy donut RD point estimates, point estimates for the first stage and the reduced form, as well as $p$-values for tests that donut trimming is negligible in the first stage and the reduced form.

We can see that donut trimming substantially changes the estimate of the first stage parameter, the jump in treatment probabilities, from 0.14 to 0.94. The $\Delta$ specification test yields a $p$-value of 0.002, indicating that this difference is unlikely to be due to random sampling alone. Our new $\Gamma$ test  yields only a $p$-value of 0.088. The estimate of the reduced form parameter, the jump in log wages, is less affected by donut trimming, and both of our specification tests yield $p$-values well above commonly used significance levels.
Taken together, the evidence from both stages provides strong evidence against the conventional fuzzy RD design in the data.

\begin{table}[!t]
	\centering
	\caption{Estimates of the effect of contract type on (log) wages}\label{tab:bau}
	\centering
	\begin{tabular}[t]{lccccc}
		\toprule
		Parameter			& $d$ &  Estimate & 95\% CI  & $\Delta$: $p$-value & $\Gamma$: $p$-value\\
		\midrule
		\multirow{2}{*}{First stage} & -
		& 0.14 & (-0.31, 0.59)  & - &  - \\
		&		 4 &  0.94 & (-0.11, 1.99) & 0.002 & 0.088\\[2ex]
		\multirow{2}{*}{Reduced form} 	& -  &
		0.22 & (-0.36, 0.79) 
		 & -  &  -\\
				& 4 &  -0.39 & (-1.62, 0.84) 
		 & 0.486 & 0.848\\[2ex]
		\multirow{2}{*}{Main parameter} & -	 & 1.52 & $(-\infty, \infty)$   & --  & --  \\
		& 4	  &  -0.41 & $(-\infty, \infty)$ & 0.005 & 0.177\\
		\bottomrule
	\end{tabular} 

	{\raggedright \footnotesize \textit{Notes:} Empirical results for the teacher-contract application of \citet{bau2020teacher}. The conventional specification uses all observations in the eight-year window around the reform; the donut specification with $d=4$ excludes the four-year low-hiring period immediately before the cutoff. The first stage refers to the jump in the probability of a temporary contract, and the reduced form refers to the jump in log monthly wages. The main parameter reports the corresponding fuzzy RD estimate with bias-aware Anderson--Rubin--type  confidence sets. The $\Delta$ and $\Gamma$ columns report $p$-values from the donut specification tests described in Section~\ref{sec::specification_testing} and Appendix~\ref{appendix:fuzzy}. We set $M_T=0.05$ and $M_Y=0.056$, where the latter value corresponds to the smoothness bound proposed by \cite{imbens2019optimized}. Estimates are based on the uniform kernel and standard errors are based on nearest-neighbor variance estimates. 
\par}
\end{table}

\subsection{Low Birthweight} We next revisit data from
\cite{almond2010estimating}, who study the effect of medical care on  infant mortality among low birthweight babies.\footnote{We obtained the data set from the NBER webpage: \url{http://data.nber.org/lbid/adkw.dta}.} Their study exploits that infants with a birthweight below 1,500 g often receive additional medical treatment. Using a conventional regression discontinuity design with a bandwidth of 85g suggests that 1-year mortality is approximately 1 percentage point lower just below the 1,500 g threshold, compared to 5.5\% just above it. The left panel of Figure~\ref{fig:almond} in the introduction shows the data together with a local linear regression fit.

\citet{barreca2011saving} argue that these results are driven by the heaping points at 1500g (rounding) and 1503g (corresponding to 53 oz). We revisit their analysis and sequentially exclude observations within $\pm d$ grams of the 1,500 g threshold, with $d$ ranging from 0 to~4. Because birth weights are recorded in integer grams, this corresponds, for example, to using a theoretical donut size of $d+0.01$. All estimates use the original bandwidth of $h = 85$ g and a uniform kernel. The right panel of Figure~\ref{fig:almond} in the introduction shows the data together with a local linear donut regression fit for the case in which $d=3$.
We set the smoothness bound to $M = 10^{-6}$, which implies that the conditional mean function $\mu^*$ can differ by at most 0.1 \% from a straight line over the 85 g window. The results show that even just excluding the observations with exactly 1500g yields a ``marginally significant'' $p$-value of 10\% from our $\Delta$ test, and when excluding observations that deviate at most 3g from the treatment threshold, the $p$-values are below commonly used significance levels.

\begin{table}[!b]
	\centering\caption{Estimates of the effect of crossing low birth weight threshold on 1-year infant mortality}\label{table::donut_birthweight}
	\begin{tabular}{@{} cccccc @{}}
		\toprule
		Parameter &	$d$ & Estimate (pp) & 95\% CI \ & $\Delta$: $p$-value & $\Gamma$: $p$-value\\
		\midrule
		\multirow{6}{*}{Main Parameter}&	-- &  0.95 & (0.46, 1.44) & -- & --\\
		&	0 & 0.54 & (0.04, 1.04) & 0.102 & --\\
		&	1 & 0.55& (0.05, 1.05) & 0.110 & --\\
		&	2 & 0.53 & (0.03, 1.04) & 0.096 & 0.659\\
		&	3 & 0.16 & (-0.47, 0.80) & 0.001 & 0.078\\
		&	4 & 0.18 & (-0.46, 0.82) & 0.001 & 0.008\\
		\bottomrule
	\end{tabular}

	{\raggedright \footnotesize  Note: Empirical results for low-birthweight application based on \citet{almond2010estimating}. The conventional specification uses a bandwidth of $h=85$g and a uniform kernel around the 1,500g threshold. Donut specifications exclude observations within $\pm d$ grams of the threshold (so $d=0$ means dropping observations with reported birth weight of exactly 1,500g). Estimates are effects on 1-year infant mortality, in percentage points. The $\Delta$ and $\Gamma$ columns report $p$-values from the donut specification tests described in Section~\ref{sec::specification_testing}. The $\Gamma$ test may not be appropriate here when only a few birthweight support points lie inside the donut. Standard errors are based on nearest-neighbor variance estimates. \par}

\end{table}

\section{Conclusions}\label{sec::conclusions}

This paper provides a theoretical foundation for donut regression discontinuity designs, which are widely used in empirical practice but typically implemented without formal guidance. Our results clarify the trade-offs introduced by donut trimming and allow for valid statistical inference in both point estimation and diagnostic settings. An important open question concerns the choice of the donut size $d$. We conjecture that if the goal is valid causal inference, this choice cannot be made in a purely data-driven way and will have to rely on some subject-matter knowledge. In a specification testing context, however, there may be scope for a procedure that uses a range of values for $d$, without assuming that a particular one is the ``correct'' one.

\clearpage
\appendix

\section{Additional Details on Robust-Bias Correction}\label{appendix:addtheory}

In this section, we present additional results that further characterize the properties of robust bias correction (RBC) methods \citep{calonico2014robust} in donut RD designs. For simplicity, we focus on the case in which the main bandwidth and the pilot bandwidth used for bias estimation coincide, in which case the common local linear RBC estimator with local quadratic bias correction is actually numerically identical to an (uncorrected) local quadratic regression estimator.

We begin by introducing more general notation. Consider local polynomial regressions of degree $p$. We denote the donut RD estimator based on a $p$th-order local polynomial, with donut size $d$ and bandwidth $h$, by $\widehat{\tau}_p(h,d)$. We further define the following kernel-dependent functionals:
\begin{align*}
	&B_{K,p}(c)
	=
	\int_c^1 J_{K,p}(u,c)K(u)u^{p+1}\,du,
	 \quad
S_{K,p}(c)
=
\int_c^1 J_{K,p}(u,c)^2K(u)^2\,du, \\ &J_{K,p}(u,c)
	=
	e_1^\top \left(\int_c^1 r_p(v)r_p(v)^\top K(v)\,dv \right) ^{-1} r_p(u),
\end{align*}
 where  $r_p(u)=(1,u,\ldots,u^p)^\top$, $c \in [0,1)$ is a constant and $e_1$ is the first unit vector of length $p+1$. 
 Let
 $b_p(h,d)
 :=
 \E[\widehat\tau_p(h,d)\mid\mathcal X_n]-\tau^*$ and $s_p^2(h,d)
 :=
 \Var(\widehat\tau_p(h,d)\mid\mathcal X_n).$ The following proposition generalizes Theorem~\ref{theorem:pointest_smalldonut}.

\begin{proposition}\label{prop:pointest_smalldonut_vp}
	Suppose that (i) Assumption~\ref{ass:sd} holds; and (ii) Assumptions~\ref{ass:rv}--\ref{ass:kernel} hold.
	Further assume that, on each side of the cutoff, $\mu^*$ is $p+1$ times continuously differentiable up to the cutoff with a uniformly bounded derivative of order $p+1$.
	Then,
	\begin{align*}&\quad\frac{\widehat\tau_{p}(h, d)-b_{p}(h, d)-\tau^*}{s_{p}(h, d)}\stackrel{d}{\to} N(0,1),\\
		&b_{p}(h, d) = h^{(p+1)} B_{K,p}(c)\frac{\mu_{+}^{\star\; (p+1)} - (-1)^{(p+1)}  \mu_{-}^{\star\; (p+1)}  }{(p+1)!}  + o_P(h^{(p+1)}),\\
		& s_{p}^2(h, d) = \frac{1}{n h}  S_{K, p}(c) \left(\frac{\sigma^2_{+}}{f_{+}} +\frac{\sigma^2_{-}}{f_{-}} \right) + o_P\left(\frac{1}{n h}\right).
	\end{align*}
\end{proposition}

\begin{figure}[tbp]
	\centering

	\setlength{\tabcolsep}{2pt}
	\begin{tabular}{cc}
		\includegraphics[width=.35\textwidth]{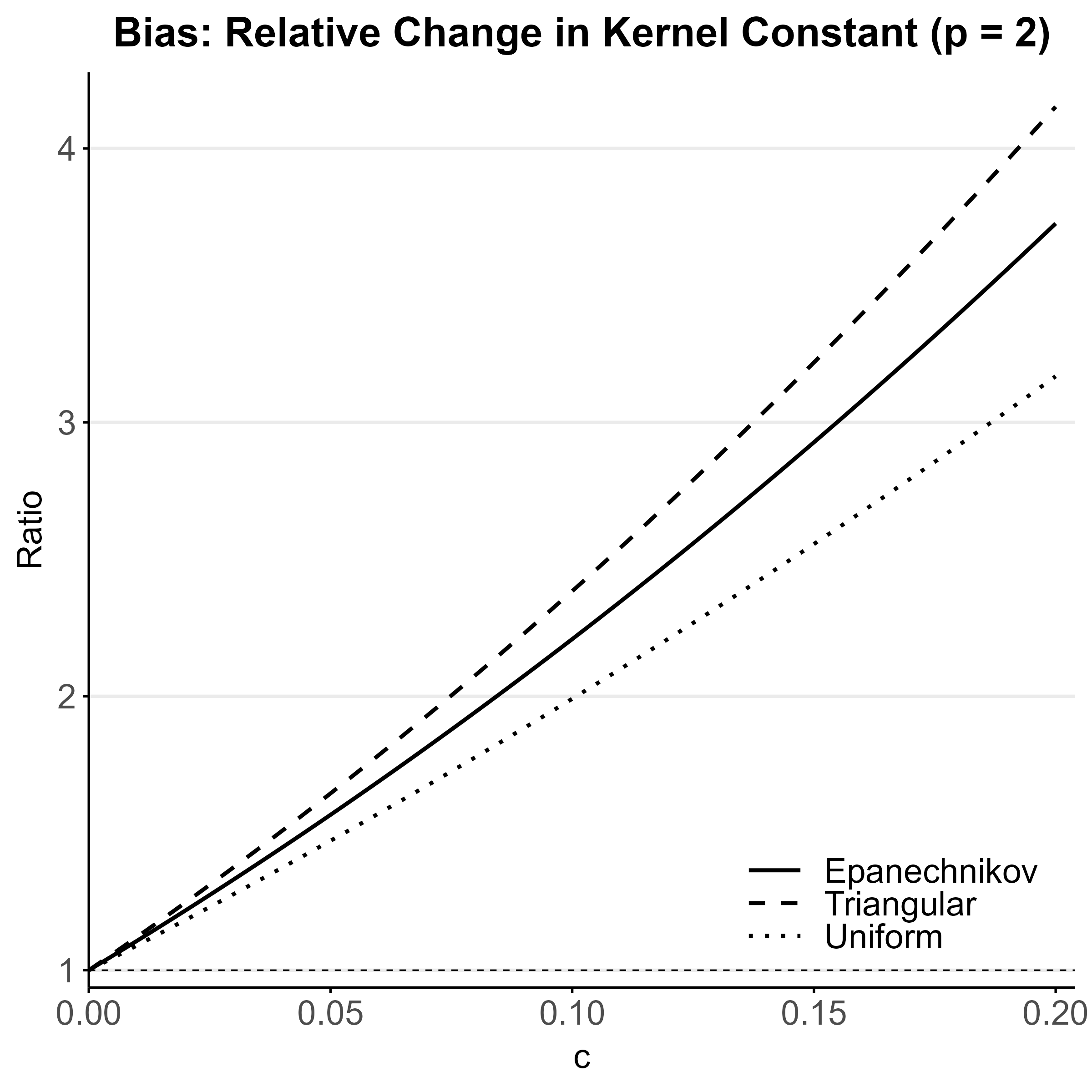} &
		\includegraphics[width=.35\textwidth]{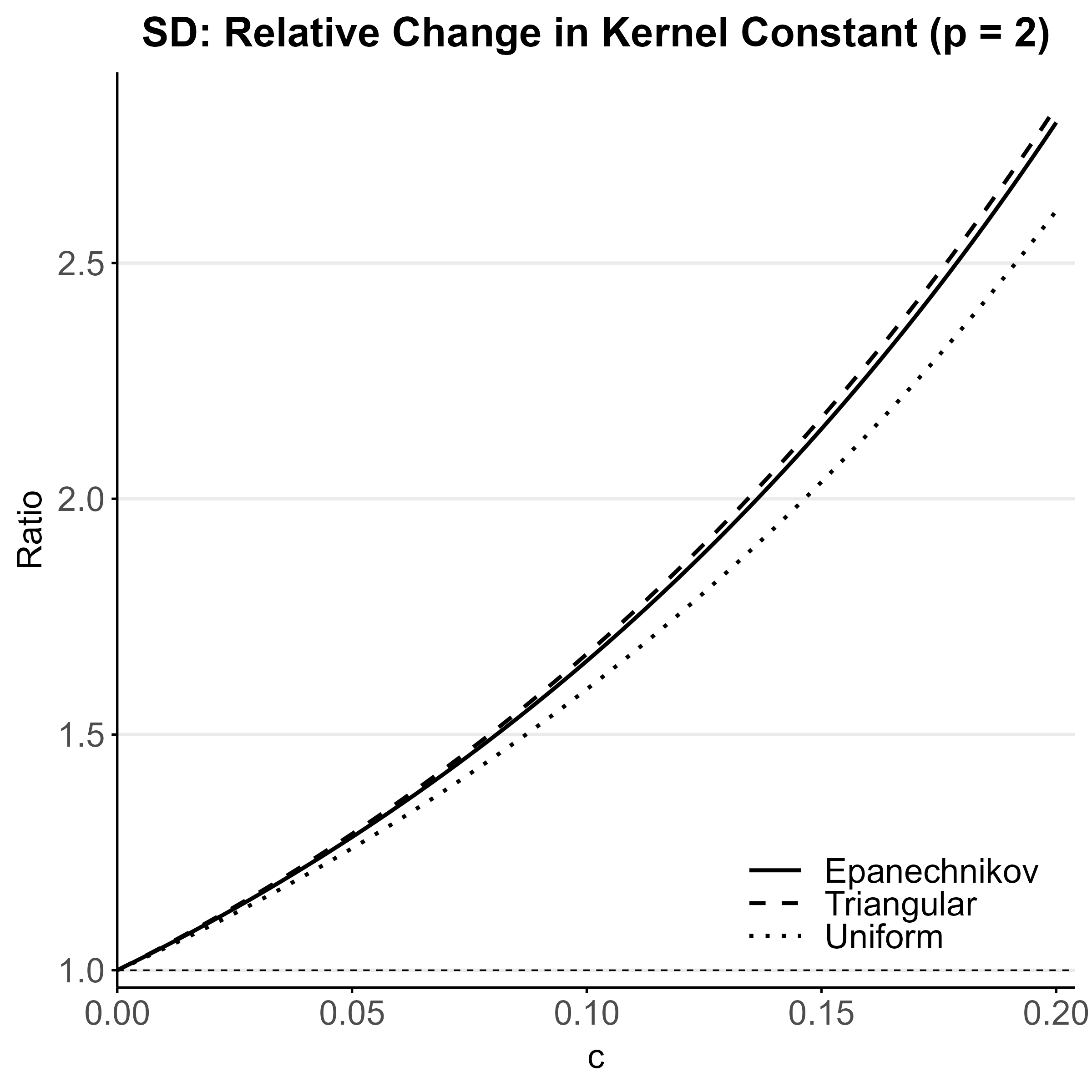} \\
		\includegraphics[width=.35\textwidth]{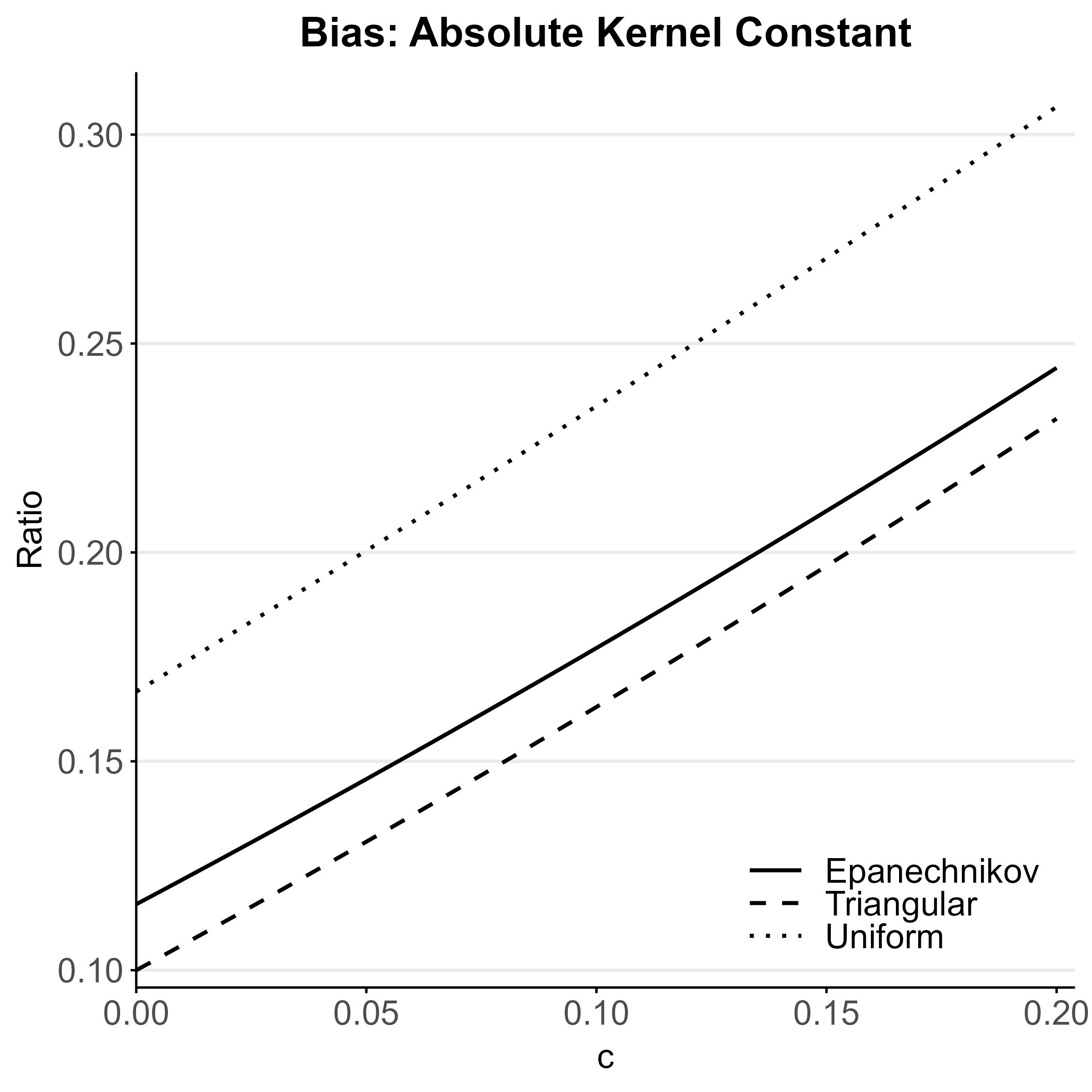} & 			
		\includegraphics[width=.35\textwidth]{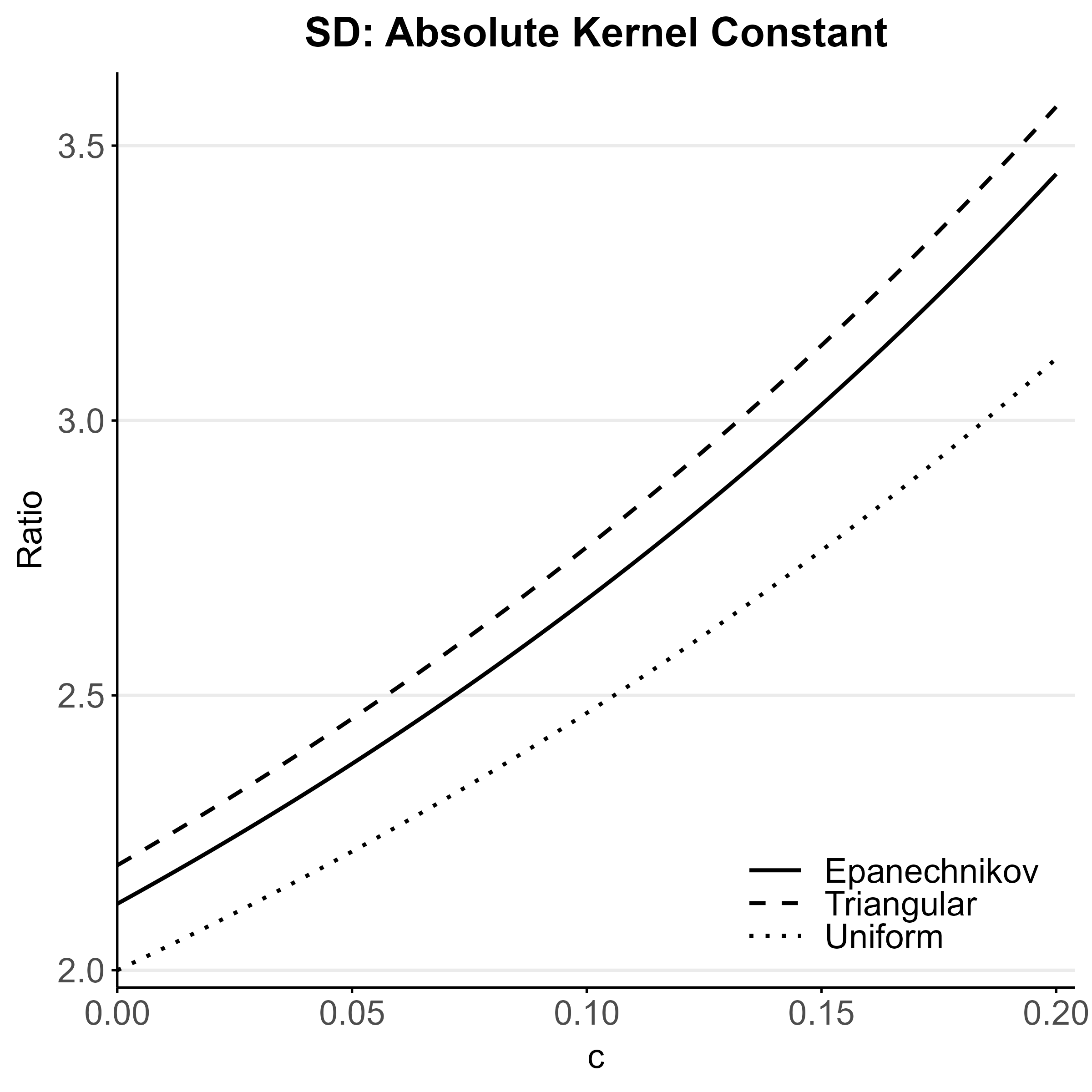}\\
		\includegraphics[width=.35\textwidth]{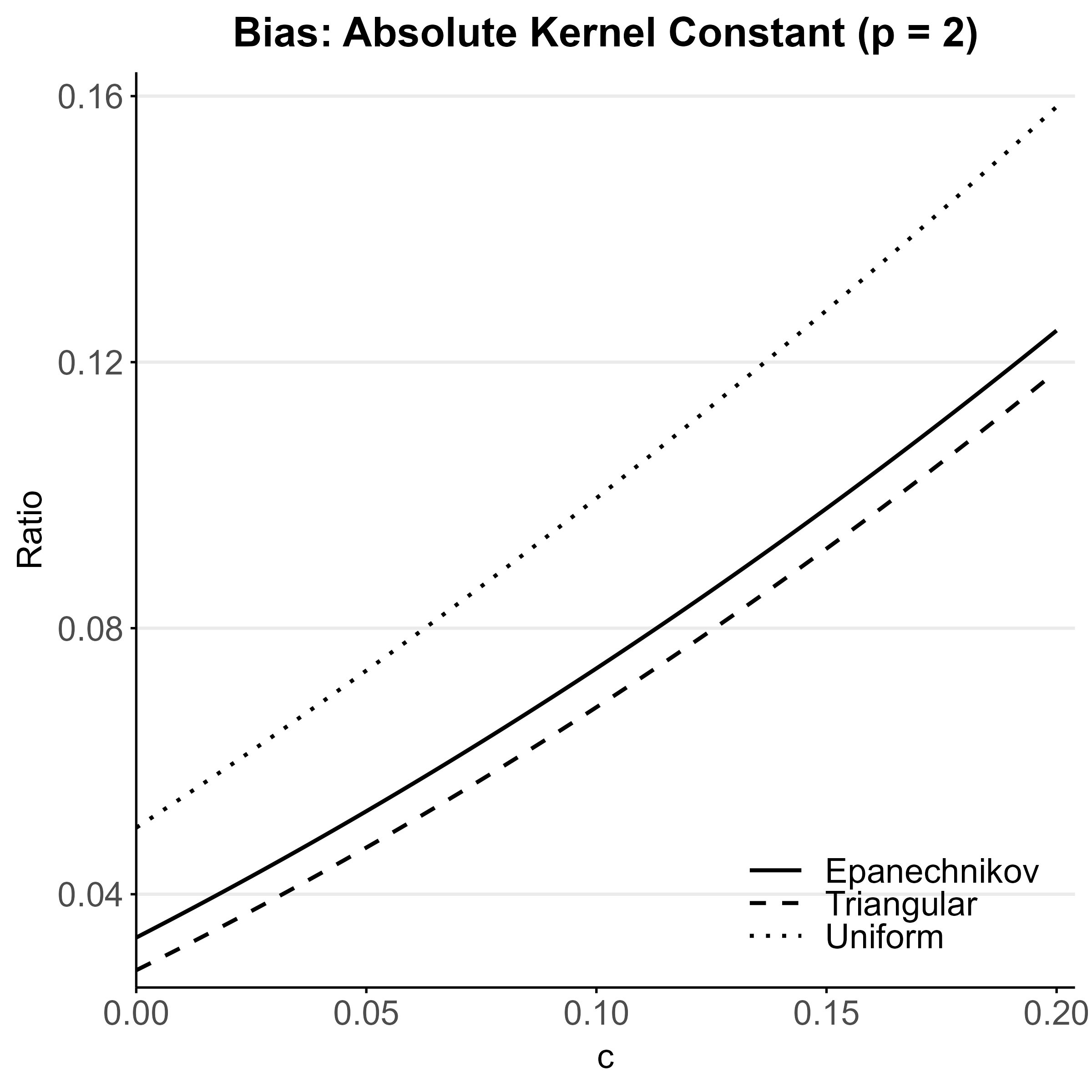} &
		\includegraphics[width=.35\textwidth]{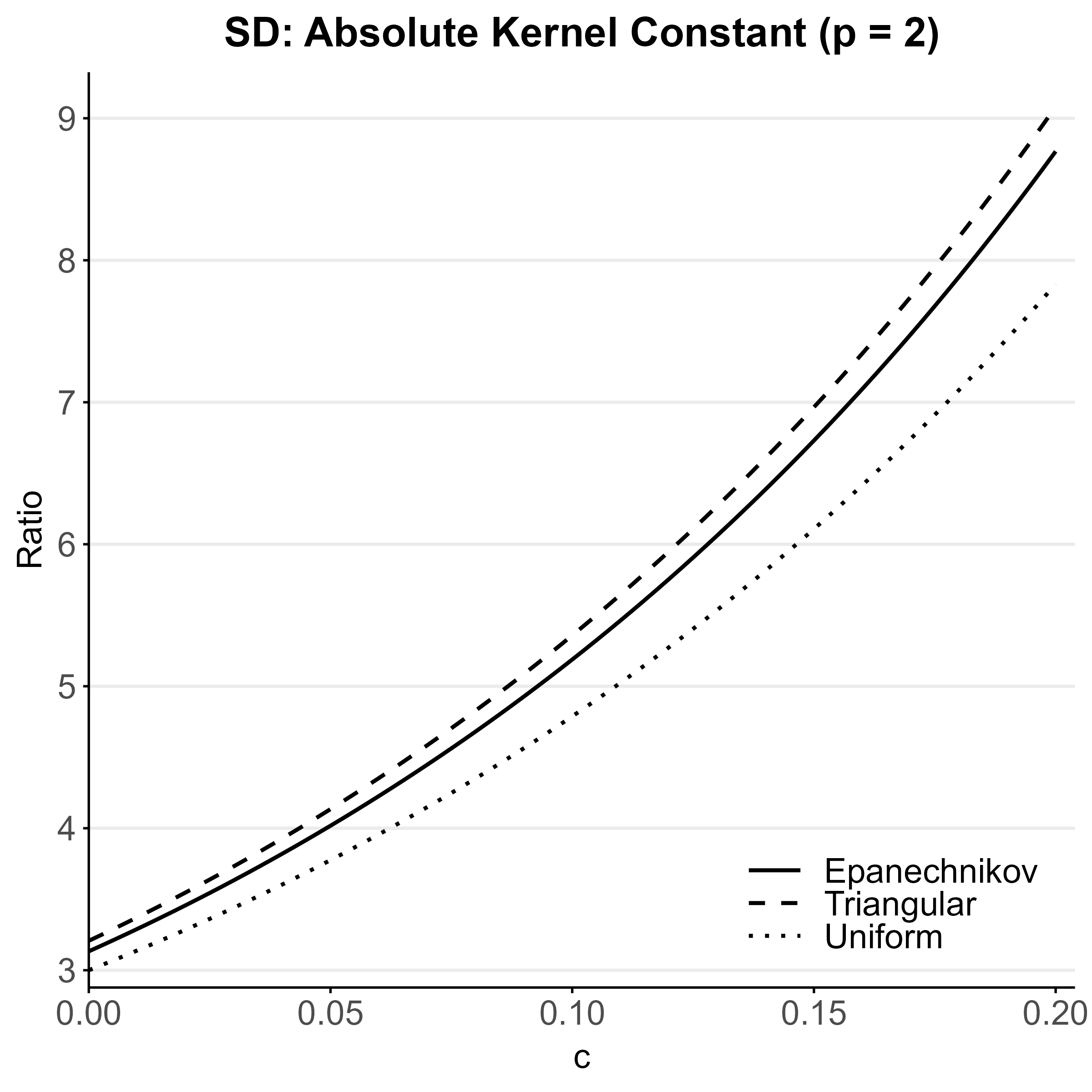} \\
	\end{tabular}

	\caption{Ratios and absolute values of kernel constants in the asymptotic bias and standard deviations of the donut and the conventional RD estimator for uniform, triangular, and Epanechnikov kernels for local linear and local quadratic regressions.}
	\label{fig:full_bias_var_const}
\end{figure}

As in the local linear case, the asymptotic bias and variance of the local polynomial RD estimator are affected by the donut size only through the kernel constants $B_{K,p}(c)$ and $S_{K,p}(c)$. 
Figure~\ref{fig:full_bias_var_const} plots the ratios $B_{K,2}(c)/B_{K,2}(0)$ and $\sqrt{S_{K,2}(c)/S_{K,2}(0)}$ for $c\in(0,.2)$ and three commonly used kernel functions. In the canonical RBC setting in which case the bandwidth is chosen to be optimal for the local linear estimator, for example with $h$ proportional to $n^{-1/5}$, the bias term is of smaller order than the standard deviation term. In this case, the figure shows that even for a moderate value such as $c=.1$, corresponding to a donut RD design that removes observations whose running-variable values differ by less than 10\% of the chosen bandwidth from the cutoff, the standard deviation increases by 60\% to 66\%, depending on the kernel function used.\footnote{
For illustration, we also plot the absolute values of $B_{K,p}(c)$ and  $S_{K,p}(c)$ for $p \in \{1,2\}$ in Figure~\ref{fig:full_bias_var_const}.}

The usual RBC confidence interval with nominal level $1-\alpha$ is then given by
$$C_n^{rbc}(d) =\left[\widehat\tau_{2}(h,d) \pm \textnormal{cv}_{1-\alpha}\left(0\right)\widehat s_2(h, d) \right].$$
It can be shown to have correct (pointwise) asymptotic coverage under small donut asymptotics and standard regularity conditions.

\begin{proposition}\label{prop:inference_smalldonut}
	Assume that the conditions of Proposition~\ref{prop:pointest_smalldonut_vp} hold for $p=2$. Further assume that $h=h_0 \cdot n^{-1/5}$ for some constant $h_0 >0$ and that $\widehat s_{2}^2(h,d) =s_{2}^2(h,d)(1+o_P(1))$. Then
	$$\lim_{n\to\infty} P(\tau^*\in C^{rbc}_n(d)) \geq 1-\alpha. $$
\end{proposition}

We can compare the  length of the donut confidence interval $C^{rbc}_n(d)$ to that of its conventional counterpart $C^{rbc}_n\equiv C^{rbc}_n(0)$ in the case where $\mu$ equals $\mu^*$. Note that in contrast to bias-aware confidence intervals, the presence of a donut  affects the length of the RBC interval only through the standard error. To obtain an analytic  result, it  suffices to consider a generic bandwidth $\bar{h}_{\textnormal{rbc}}\propto n^{-1/5}$. The presence of a donut then changes the length of the confidence interval by a factor of
\begin{align*}
	\frac{\widehat s_2(\bar{h}_{\textnormal{rbc}}, d)}{ \widehat{s}_2(\bar{h}_{\textnormal{rbc}},0)}=\sqrt{\frac{S_{K,2}(c)}{S_{K,2}(0)}} + o_P(1). \label{eq:rbc.asy.length.increase}
\end{align*}
The leading factor is plotted in Figure~\ref{fig:full_bias_var_const} as discussed above. For $c=.1$, the presence of a donut thus increases the length of the RBC confidence interval   by 60\% to 66\%, depending on the kernel function used. Recall  from the main body of the paper that a bias-aware confidence interval increases in length by only 26\% to 33\% for the same donut size. The power of RBC inference is thus more severely affected by the presence of a donut.

Following the arguments in Section~\ref{sec::specification_testing}, it is straightforward to construct robust bias-corrected specification tests for both the donut and the no-donut cases under the small-donut asymptotic framework; we omit the details for brevity.

\section{Proofs of main results}\label{appendix:proofs}

In this section, we collect the proofs of the main theoretical results in this paper. Note that some of our extensions involve the use of higher-order local polynomials. We therefore formulate many intermediate results directly in terms of higher-order local polynomials.

\subsection{Additional Notation}\label{A:sec:weights}

To simplify the exposition, we use the following notation: We let  $I^+=\{i \in [n]: X_i \geq 0\}$,  $I^-=\{i \in [n]: X_i < 0\}.$  We define  estimators of the jump of the conditional expectation of the outcome  $Y_i$ at the cutoff using the $p$-th order local polynomial regression as
\begin{align*}
\widehat\tau_p(h,d) &=\sum_{i \in [n]} w_{i,p}(h,d)Y_i, \text{where } w_{i,p}(h,d)=w_{i,p}^+(h,d)-w_{i,p}^-(h,d) \\
w_{i,p}(h,d)&=e_1^\top\Big(\sum_{j=1}^n K(X_j/h)\1{|X_j|\ge d}\widetilde X_{p,j}\widetilde X_{p,j}^\top\Big)^{-1}K(X_i/h)\1{|X_i|\ge d}\widetilde X_{p,i},\\
\text{so that } &   
 w_{i,p}^+(h,d)=\1{X_i\ge0}w_{i,p}(h,d),\text{and } w_{i,p}^-(h,d)=-\1{X_i<0}w_{i,p}(h,d),
\end{align*}
where $\widetilde{X}_{p,i}=( \1{X_i \geq 0} , 1, X_i, X_i \1{X_i \geq 0}, \ldots , X_i^p,  \1{X_i \geq 0}X_i^p)^\top$, $K_h(v)=K(v/h)/h$, $K^+_h(v)=K_h(v)\1{v\geq 0}$,  $K^-_h(v)=K_h(v)\1{v<0}$.
Note that in the special case $p=1$ and for $\star\in\{+,-\}$,
the local linear weights can be further expressed as
\begin{align*}
	w^\star_{i,1}(h,d)  = \frac{ \frac{1}{n} K^\star_h(X_i) \1{|X_i| \geq d} \left(\mathcal  V^\star_{2}(h,d)   - \mathcal  V^\star_{1}(h,d) (X_i/h) \right)  }{ \mathcal V^\star_{2}(h,d) \mathcal V^\star_{0}(h,d)- (\mathcal  V^\star_{1}(h,d))^2 },
\end{align*}
$$\text{with } \mathcal V^\star_{l}(h,d) = \frac{1}{n} \sum_{i \in [n]} K^\star_h(X_i) \1{|X_i| \geq d} (X_{i}/h)^l$$
We also let $[a]_+=\max\{a,0\}$ and  $[a]_-=\min\{a,0\}$ for $a \in \mathbb R$ and  we define
\begin{align*}
	w_{i, \Delta}(h,d)&=	w_{i, 1, \Delta}(h,d)= w_{i}(h,d)-w_{i}(h,0)\\
	w_{i, \Gamma}(h,d)	& = w_{i, 1, \Gamma}(h,d) =w_{i}(h,d)-w_{i}(d,0),
\end{align*}
where we will sometimes omit the p index, as it is used only for $p=1$.

\subsection{Auxiliary Results}
Before proving the main results, we state a generic auxiliary result for the
weighted estimators considered below. Let
$
\diamond\in\{\mathrm{RD},\Delta,\Gamma\}.
$
Here $\mathrm{RD}$ refers to the baseline donut RD estimator, while $\Delta$ and
$\Gamma$ refer to the two specification-test statistics under their respective null
hypotheses. We attach the subscript $\diamond$ to all corresponding quantities:
the estimator $\widehat\tau_{p,\diamond}$, weights $w_{i,p,\diamond}(h,d)$,
standard errors $s_{p,\diamond}(h,d)$ and $\widehat s_{p,\diamond}(h,d)$, bias
$b_{p,\diamond}(h,d)$, and worst-case bias bound $\bar b_{p,\diamond}(h,d)$.
The target is denoted by $\tau^*_{p,\diamond}$, with
$\tau^*_{p,\mathrm{RD}}=\tau^*$ and
$\tau^*_{p,\Delta}=\tau^*_{p,\Gamma}=0$ under the corresponding null hypotheses.

\begin{lemma}\label{lemma::clt}
	Suppose that (i)  Assumption~\ref{ass:sd} or~\ref{ass:ld} holds; (ii) Assumptions~\ref{ass:rv}--\ref{ass:kernel} hold; and (iii) $\widehat s_{p,\diamond}^2(h,d) =s_{p,\diamond}^2(h,d)(1+o_P(1))$ uniformly over $\mathcal F(M)$. Then, 
	\begin{enumerate}
		\item[(1)] $ \displaystyle \max_{i \in [n]} |w_{i,p, \diamond}(h,d)|/\sqrt{\sum_{j=1}^n w_{j,p, \diamond}(h,d)^2} = o_P(1), $
	 \item[(2)] $\displaystyle  (\wh \tau_{p,\diamond} - \mathbb E[\wh \tau_{p,\diamond}| \mathcal X_n]  ) / \wh s_{p,\diamond}(h,d) \stackrel{d}{\to} \mathcal N(0,1),$
	\item[(3)] $\displaystyle \liminf_{n\to\infty}\inf_{\mu^*\in\mathcal{F}(M)} P(\tau^*_{p,\diamond} \in  [ \wh \tau_{p,\diamond} \pm  cv_{1-\alpha}(\bar b_{p,\diamond}(h,d)/ \wh s_{p,\diamond}(h,d))  \wh s_{p,\diamond}(h,d)]) \geq 1-\alpha. $
	\end{enumerate}
\end{lemma}
\begin{proof}
	Part~(1) follows from standard kernel calculations.

To show part (2), let $\varepsilon_i=Y_i-\mu(X_i)$. By the definition of
	$b_{p,\diamond}(h,d)$,
	\[
	\widehat\tau_{p,\diamond}
	-
	\mathbb E[\wh \tau_{p,\diamond}| \mathcal X_n]
	=
	\sum_{i \in [n]} w_{i,p,\diamond}(h,d)\varepsilon_i .
	\]
	Conditional on $\mathcal X_n$, the summands are independent and have mean zero. Moreover,
	by Assumption~\ref{ass:moments}, for some $\delta\in(0,q-2]$,
	\[
	\sum_{i \in [n]}
	\E\!\left[
	\left|
	\frac{w_{i,p,\diamond}(h,d)\varepsilon_i}
	{s_{p,\diamond}(h,d)}
	\right|^{2+\delta}
	\middle|\mathcal  X_n
	\right]
	\leq
	C
	\left(
	\frac{\max_i |w_{i,p,\diamond}(h,d)|}
	{\sqrt{\sum_j w_{j,p,\diamond}(h,d)^2}}
	\right)^\delta
	=
	o_P(1),
	\]
	where we use that the conditional variances are bounded away from zero and
	above. Hence the conditional Lyapunov triangular-array CLT implies
	\[
	\frac{\sum_i w_{i,p,\diamond}(h,d)\varepsilon_i}
	{s_{p,\diamond}(h,d)}
	\overset{d}{\to}N(0,1).
	\]
	The result with $\widehat s_{p,\diamond}(h,d)$ follows from
	$\widehat s_{p,\diamond}^2(h,d)/s_{p,\diamond}^2(h,d)\to_P1$ and Slutsky's theorem.

	The preceding Lyapunov argument is uniform over $\mathcal F(M)$, and together with the uniform consistency of $\widehat s_{p,\diamond}(h,d)$, part (3) follows from part (2) and the definition of $\bar b_{p,\diamond}(h,d)$.
	\end{proof}

\subsection{Proof of Theorem~\ref{theorem:pointest_smalldonut}}
Theorem~\ref{theorem:pointest_smalldonut} follows from Proposition~\ref{prop:pointest_smalldonut_vp}
by setting $p=1$. \qed

\subsection{Proof of Proposition~\ref{prop:bandwidth}}

	To simplify notation, write
	\[
	T_1=\left(\frac{\mu_+''-\mu_-''}{2}\right)^2,\qquad
	T_2=\frac{\sigma_+^2}{f_+}+\frac{\sigma_-^2}{f_-},
	\]
	so that
	\[
	h_{MSE}(d)
	=
	\arg\min_{h>d}
	\left\{
	h^4B_K(d/h)^2T_1+\frac{1}{nh}S_K(d/h)T_2
	\right\},
	\]
	where $T_1>0$ and $T_2>0$. For $d=d_0n^{-1/5}$ with $d_0>0$ fixed, define $c=d/h\in(0,1)$, so that
	$h=d/c=(d_0/c)n^{-1/5}.$ Now let
	\[R(c;d_0)=
	\left(\frac{d_0}{c}\right)^4B_K(c)^2T_1
	+
	\frac{c}{d_0}S_K(c)T_2.
	\]
	Then minimizing the original objective function over $h>d$ is
	equivalent to minimizing $R(c;d_0)$ over $c\in(0,1)$. In particular, if
	\[
	c^*(d_0)=\argmin_{c\in(0,1)}R(c;d_0)
	\]
	exists and is unique, then
	\[
	h_{MSE}(d)=\frac{d_0}{c^*(d_0)}n^{-1/5}.
	\]

	We next establish existence and uniqueness of a minimizer.
	Under Assumption~\ref{ass:kernel}, the truncated kernel moments
	$\int_c^1 u^jK(u)\,du$ and $\int_c^1 u^jK(u)^2\,du$
	are continuously differentiable on $(0,1)$.
	It follows  that $B_K(\cdot)$ and $S_K(\cdot)$ are continuously differentiable on $(0,1)$. Since $K$ is bounded with support contained in $[-1,1]$, both functions are continuous on $[0,1)$. Moreover, $S_K(c)>0$ for all $c\in[0,1)$, and $S_K(c)\to\infty$ as $c\uparrow 1$.
	
	Since $B_K(\cdot)$ and $S_K(\cdot)$ are continuous on $(0,1)$, so is $R(\cdot;d_0)$. Moreover, because $B_K(0)\neq 0$,
	\[
	R(c;d_0)\geq \left(\frac{d_0}{c}\right)^4B_K(c)^2T_1\to\infty
	\text{ as }c\downarrow 0.
	\]
	Likewise,
	\[
	R(c;d_0)\geq \frac{c}{d_0}S_K(c)T_2\to\infty
	\text{ as }c\uparrow 1,
	\]
	By continuity, $R(\cdot;d_0)$ then attains its global minimum $c^*(d_0)$ in the interior. The global minimum therefore satisfies the first-order condition
	\[
	\frac{\partial R(c^*(d_0);d_0)}{\partial c}=0.
	\]
	Differentiating $R(c;d_0)$ gives
	\[
	\frac{\partial R(c;d_0)}{\partial c}
	=
	d_0^4T_1\frac{2B_K(c)\bigl(cB_K'(c)-2B_K(c)\bigr)}{c^5}
	+
	\frac{T_2}{d_0}\bigl(S_K(c)+cS_K'(c)\bigr).
	\]
	Hence the first-order condition can be written as
	\[
	d_0^5\frac{T_1}{T_2}=c^5G_K(c),
	\]
	where $G_K(c)$ is as defined in the statement of the Proposition. The function $c^5G_K(c)$ is continuous and strictly increasing on $(0,1)$, with limits $0$ and $\infty$ at the two boundaries. It follows that, for every $d_0>0$, the first-order equation
	has a unique solution $c\in(0,1)$. Since every interior minimizer must satisfy this equation, the minimizer $c^*(d_0)$ is unique.

	To prove monotonicity of $d_0/c^*(d_0)$, define
	\[
	\ell_K(c)=
	\left(
	\frac{T_2}{T_1}G_K(c)
	\right)^{1/5},
	\qquad
	m_K(c)=c\,\ell_K(c)=
	\left(
	\frac{T_2}{T_1}c^5G_K(c)
	\right)^{1/5}.
	\]
	Then the first-order condition implies
	\[
	d_0=m_K(c^*(d_0)),
	\qquad
	\frac{d_0}{c^*(d_0)}=\ell_K(c^*(d_0)).
	\]
	Since $G_K(\cdot)$ is strictly positive and strictly increasing, both $m_K(\cdot)$ and $\ell_K(\cdot)$ are strictly increasing on $(0,1)$. Therefore $c^*(d_0)$ is strictly increasing in $d_0$, and so is
	\[
	\frac{d_0}{c^*(d_0)}=\ell_K(c^*(d_0)).
	\]
	This proves the monotonicity claim.

	It remains to establish the limit as $d_0\to 0$. Introduce the reparametrization $\lambda=d_0/c$,
	so that $c=d_0/\lambda.$ Then
	\[
	R(c;d_0)
	=
	\lambda^4B_K(d_0/\lambda)^2T_1+\lambda^{-1}S_K(d_0/\lambda)T_2.
	\]
	For every fixed $\lambda>0$, continuity of $B_K(\cdot)$ and $S_K(\cdot)$ at zero implies
	\[
	\lambda^4B_K(d_0/\lambda)^2T_1+\lambda^{-1}S_K(d_0/\lambda)T_2
	\to
	H_K(\lambda)\equiv\lambda^4B_K(0)^2T_1+\lambda^{-1}S_K(0)T_2
	\]
	as $d_0\to 0$. The convergence is uniform on compact subsets of $(0,\infty)$. Since
	\[
	H_K(\lambda)\to\infty
	\qquad\text{as }\lambda\downarrow 0
	\quad\text{and}\quad
	H_K(\lambda)\to\infty
	\qquad\text{as }\lambda\uparrow\infty,
	\]
	the function $H_K$ has a unique minimizer $\lambda_0$, characterized by
	\[
	4\lambda_0^3B_K(0)^2T_1=\lambda_0^{-2}S_K(0)T_2,
	\]
	or equivalently,
	\[
	\lambda_0^5=\frac{S_K(0)T_2}{4B_K(0)^2T_1}
	=
	\frac{
	S_K(0)\left(\frac{\sigma_+^2}{f_+}+\frac{\sigma_-^2}{f_-}\right)
	}{
	B_K(0)^2(\mu_+''-\mu_-'')^2
	}.
	\]
	A standard argmin continuity argument then gives
	\[
	\frac{d_0}{c^*(d_0)}\to \lambda_0
	\qquad\text{as }d_0\to 0.
	\]
	Therefore
	\[
	\lim_{d_0\to 0}h_{MSE}(d)
	=
	\lambda_0n^{-1/5}
	=
	\left(
	\frac{
	S_K(0)\left(\frac{\sigma_+^2}{f_+}+\frac{\sigma_-^2}{f_-}\right)
	}{
	B_K(0)^2(\mu_+''-\mu_-'')^2
	}
	\right)^{1/5}
	\times n^{-1/5},
	\]
	which completes the proof. \qed
	
	\bigskip

	We complete this subsection by verifying that the high-level condition imposed in
	Proposition~\ref{prop:bandwidth} holds for the uniform, triangular, and Epanechnikov
	kernels.  
	It is convenient to define
	\[
	G_K(c)=
	\frac{S_K(c)+cS_K'(c)}
	{2B_K(c)\bigl(2B_K(c)-cB_K'(c)\bigr)}.
	\]
	The calculations below show that $G_K(c)>0$, that $G_K'(c)>0$, and that $c^5G_K(c)$
	has the required boundary behavior.
    
    \medskip

	\noindent \textbf{Uniform Kernel.} For the uniform kernel, we have
	\[
	B_K(c)=-\frac{1+4c+c^2}{6},\qquad
	S_K(c)=\frac{4(1+c+c^2)}{(1-c)^3},
	\]
	so that
	\[
	2B_K(c)-cB_K'(c)=-\frac{1+2c}{3},
	\qquad
	S_K(c)+cS_K'(c)=\frac{4(1+2c)^2}{(1-c)^4},
	\]
	and therefore
	\[
	G_K(c)=\frac{36(1+2c)}{(1-c)^4(1+4c+c^2)}.
	\]
	In particular, $G_K(c)>0$ for all $c\in(0,1)$. Moreover,
	\[
	\frac{d}{dc}\log G_K(c)
	=
	\frac{2}{1+2c}+\frac{4}{1-c}-\frac{2(c+2)}{1+4c+c^2}
	=
	\frac{2(5c^3+18c^2+12c+1)}
	{(1-c)(1+2c)(1+4c+c^2)}.
	\]
	The denominator is strictly positive on $(0,1)$, and the numerator polynomial has strictly
	positive coefficients. Hence $(d/dc)\log G_K(c)>0$ on $(0,1)$, so that $G_K(c)$ is strictly
	increasing. The displayed formula also implies that $c^5G_K(c)\to0$ as $c\downarrow0$ and
	$c^5G_K(c)\to\infty$ as $c\uparrow1$.
	
	\medskip

	\noindent \textbf{Triangular Kernel.} For the triangular kernel, we have
	\[
	B_K(c)=-\frac{1+6c+3c^2}{10},\qquad
	S_K(c)=\frac{12(2+3c+3c^2)}{5(1-c)^3},
	\]
	so that
	\[
	2B_K(c)-cB_K'(c)=-\frac{1+3c}{5},
	\qquad
	S_K(c)+cS_K'(c)=\frac{24(1+3c)(1+2c)}{5(1-c)^4},
	\]
	and therefore
	\[
	G_K(c)=\frac{120(1+2c)}{(1-c)^4(1+6c+3c^2)}.
	\]
	Again $G_K(c)>0$ for all $c\in(0,1)$. Moreover,
	\[
	\frac{d}{dc}\log G_K(c)
	=
	\frac{2}{1+2c}+\frac{4}{1-c}-\frac{6(1+c)}{1+6c+3c^2}
	=
	\frac{30c(1+c)^2}{(1-c)(1+2c)(1+6c+3c^2)},
	\]
	which is strictly positive on $(0,1)$. Hence $G_K(c)$ is strictly increasing. The displayed
	formula also implies that $c^5G_K(c)\to0$ as $c\downarrow0$ and $c^5G_K(c)\to\infty$ as
	$c\uparrow1$.
	
	\medskip

	\noindent \textbf{Epanechnikov Kernel.} For the Epanechnikov kernel, the calculations become more tedious. We define the following three polynomials:
	\begin{align*}
	P_1(c)&=6c^4+36c^3+81c^2+66c+11,\\	
	P_2(c)&=129c^4+684c^3+1254c^2+924c+209,\\
		P_3(c)&=2691c^7+23418c^6+87126c^5+175620c^4+202815c^3+131394c^2+43312c+5624.
	\end{align*}
	A direct calculation gives
	\[
	B_K(c)=-\frac{P_1(c)}{5(3c^2+18c+19)},
	\]
	\[
	S_K(c)=
	\frac{48(99c^6+891c^5+3345c^4+6345c^3+6600c^2+3936c+1184)}
	{35(1-c)^3(3c^2+18c+19)^2},
	\]
	and
	\[
	G_K(c)=\frac{240\,P_3(c)}{7(1-c)^4P_1(c)P_2(c)}.
	\]
	Since all coefficients of $P_1$, $P_2$, and $P_3$ are strictly positive, we have $P_1(c)>0$,
	$P_2(c)>0$, and $P_3(c)>0$ for all $c\in[0,1)$, and hence $G_K(c)>0$ on $(0,1)$. In
	addition,
	\[
	\frac{d}{dc}\log G_K(c)
	=
	\frac{15(1+c)P_4(c)}{(1-c)P_1(c)P_2(c)P_3(c)},
	\]
	where
	\begin{align*}
		P_4(c)=&\;694278c^{14}+12694644c^{13}+107571123c^{12}+557423910c^{11}\\
		&+1963602405c^{10}+4946052924c^9+9127765662c^8+12445500924c^7\\
		&+12490270320c^6+9084915900c^5+4648824623c^4+1590174894c^3\\
		&+332171653c^2+35233220c+1103520.
	\end{align*}
	All coefficients of $P_4$ are strictly positive, so $P_4(c)>0$ for $c\ge 0$. Since
	the denominator is also strictly positive on $(0,1)$, it follows that $(d/dc)\log G_K(c)>0$
	on $(0,1)$. Hence $G_K(c)$ is strictly increasing. The displayed formula also implies that
	$c^5G_K(c)\to0$ as $c\downarrow0$ and $c^5G_K(c)\to\infty$ as $c\uparrow1$.

\subsection{Proof of Theorem~\ref{theorem:uniform_donut_estimator}}
	We first note that $$ \sup_{\mu^* \in \mathcal F(M)} |b(h,d)| \leq -  \frac{M}{2}\sum_{i \in [n]} w_{i,1}(h,d) X_i^2 \sgn(X_i),$$
	which directly follows from the  results of \cite{armstrong2018simple}. Theorem~\ref{theorem:uniform_donut_estimator} then directly follows from  Lemma~\ref{lemma::clt}.

\subsection{Proofs of  Theorems~\ref{theorem:standard_donut_test}~and~\ref{theorem:alternative_donut_test}}
	The proofs of Theorems~\ref{theorem:standard_donut_test}~and~\ref{theorem:alternative_donut_test} follow the same line of arguments and we therefore prove them jointly here. Let $\diamond \in \{\Gamma, \Delta\}$.  In the following, suppose that $n$ is sufficiently large that the weighted local polynomial regressions defining
	$\widehat\tau_{p,\diamond}$ are well defined.  If
	\begin{align}\label{equ::delta_maxbias}
		\sup_{\mu \in \mathcal F(M)} |b_\diamond(\mu, h)| \leq -  \frac{M}{2}\sum_{i \in [n]} w_{i,1,\diamond}(h,d) X_i^2 \sgn(X_i),
	\end{align}
	then Theorems~\ref{theorem:standard_donut_test}~and~\ref{theorem:alternative_donut_test} directly follow from Lemma~\ref{lemma::clt}.
	To show~\eqref{equ::delta_maxbias}, let
	$$b_\diamond(\mu, h) = \sum_{i \in I^+} w_{i,1,\diamond}(h,d) (\mu(X_i) - \mu(0^+)) + \sum_{i \in I^-} w_{i,1,\diamond}(h,d) (\mu(X_i) - \mu(0^-))  \equiv L_{\diamond,+} + L_{\diamond,-}.  $$
	Consider the term 	$L_{\diamond,+}.$  By the properties of weights of local linear regressions and by using a second order Taylor expansion and basic properties of local linear regression weights, e.g., $\sum_{i \in I^+} w_{i,1}(h,d)=1 $ and $\sum_{i \in I^+} w_{i,1}(h,d)X_i=0$, it holds that
	\begin{align*}
		L_{\diamond,+} & =\sum_{i \in I^+} w_{i,1,\diamond}(h,d)  \int_{0}^{X_i} \mu^{''}(s)  (X_i-s)  ds
		=  \int_{0}^\infty \mu^{''}(s) \sum_{i: X_i >s}   w_{i,1,\diamond}(h,d) (X_i-s)  ds \\
		& \equiv \int_{0}^\infty \mu^{''}(s) v_\Diamond(s,h,d)  ds,
	\end{align*}
	where  we use Fubini's Theorem and note that
	$$
	v_{\diamond}(s, h, d) := \sum_{i: X_i > s} w_{i,1,\diamond}(h,d) (X_i - s) = \sum_{i \in I_+: X_i \leq s}   w_{i,1,\diamond}(h,d) (s-X_i)$$
	as $\sum_{i \in I^+} w_{i,1,\diamond}(h,d) = \sum_{i \in I^+} w_{i,1,\diamond}(h,d) X_i=0$.
	Now, if  
	\begin{equation}
		v_{\diamond}(s, h, d) \leq 0 \quad \text{for all } s \geq 0 \text{ and } \diamond \in \{\Delta, \Gamma\}, \label{eq::vshd}
	\end{equation}
	then
	\[
	|L_{\diamond,+}(h,d)|
	\leq
	-M\int_0^\infty v_\diamond(s,h,d)\,ds
	=
	-\frac{M}{2}
	\sum_{i\in I^+}
	w_{i,1,\diamond}(h,d)X_i^2.
	\]
	Applying the same argument to the left side of the cutoff yields
	\[
	|L_{\diamond,-}(h,d)|
	\leq
	\frac{M}{2}
	\sum_{i\in I^-}
	w_{i,1,\diamond}(h,d)X_i^2,
	\]
which shows that inequality~\eqref{equ::delta_maxbias} holds.

	\medskip

	To show~\eqref{eq::vshd}, we first note that the function $v_\diamond(s,h,d)$ is continuous for all $s\geq 0$ and we further note that $v_\diamond(s,h,d)$ is differentiable for all $s \notin   \{X_i\}_{i \in I^+}$.
	The proof is then completed by the following case distinctions.

\subsection*{Bias bound for  $\Gamma$}
	We analyze  $v_{\Gamma}(s, h, d)$ 	in two cases:

	\noindent\textbf{\noindent Case 1: \(0\leq s < d\).}
	In this case, and for $X_i \leq s$,  the weights reduce to:
	$
	w_{i,1, \Gamma}(h,d) = - w_{i,1}(d,0).$
	Therefore,
	\[
	v_{\Gamma}(s,h,d) = \sum_{i  \in I_+: X_i \leq s} w_{i,1}(d,0)( X_i-s),
	\]
	which equals the intercept of a weighted local linear regression of \([X_i - s]_-\) on \(X_i\), for \(X_i \geq 0\). This intercept is  negative by properties of local linear regression under standard kernels. Thus, \(v_{\Gamma}(s,h,d) \leq 0\) for \(s < d\).

	\smallskip
	\noindent\textbf{\noindent Case 2: \(s \geq d\).}
	Now  the expression becomes:
	\[
	v_{\Gamma}(s,h,d) = \sum_{i \in I_+: X_i \leq s} w_{i,1}(h,d)(s - X_i) - s.
	\]

	The first part of this expression  is the intercept from a weighted linear regression of \([s - X_i]_+\) on \(X_i\) for \(X_i > d\). As the intercept of this regression is always less than or equal to \(s\), the entire expression is again non-positive,
	$v_{\Gamma}(s,h,d) \leq 0.$

\subsection*{Bias bound for  $\Delta$}

	We analyze  $v_{\Delta}(s, h, d)$
	in three cases:

	\noindent\textbf{\noindent Case 1: \boldmath\(s \geq h\).}
	For all \(i\), if \(X_i > s \geq h\), then \(w_{i,1}(h,d) = w_{i,1}(h,0)   = 0\) due to kernel support. Thus,  it trivially holds that
	$
	v_\Delta(s, h, d) = \sum_{i: X_i > s} 0 \cdot (X_i - s) = 0.
	$

	\smallskip

	\noindent\textbf{\noindent Case 2: \boldmath\(s < d\).}
	For all \(i\) such that \(X_i \leq s\), the weights \(w_{i,1}(h,d) = 0\), and thus,
	\[
	v_\Delta(s,h,d) = \sum_{i \in I^+: X_i \leq s} w_{i,1,\Delta}(h,d) (s-X_i)=  -\sum_{i \in I^+: X_i \leq s} w_{i,1}(h,0)(s-X_i).
	\]
	
	This expression is the negative of the intercept of a weighted local linear regression of \([s - X_i]_+\) on \(X_i\). As this intercept is nonnegative, \(v_\Delta(s,h,d) \leq  0\).

	\smallskip

	\noindent\textbf{\noindent Case 3: \boldmath$s \in (d,h)$.}
	Note that we have shown that \(v_\Delta(d,h,d) \leq  0\) and \(v_\Delta(h,h,d) = 0\). In the following we show that $\frac{\partial}{\partial s}  v_\Delta(d,h,d) \leq 0$ and that $\frac{\partial}{\partial s}  v_\Delta(s,h,d)$ changes its sign at most once on $(d,h)$, from nonpositive to nonnegative, and thus \(v_\Delta(s,h,d) \leq 0\) for all $s\geq0$.
	Note that $v_\Delta(s, h, d)$ is a continuous function and  that for $s \notin \{X_i\}_{i=1}^n$,
	\[
	\frac{\partial}{\partial s} v_\Delta(s,h,d) =\sum_{i \in I^+: X_i \leq s} w_{i,1,\Delta}(h,d)
	= A(s) - T_2,
	\]
	\[
	A(s) := \sum_{i:\, d \le X_i \le s} K_h^+(X_i)\,\varrho(X_i),
	\qquad
	T_2 := \sum_{i \in I_+:\, X_i < d} w_{i,1}(h,0),
	\]
	where $T_2 \ge 0$ is constant in $s$ and
	\[
	\varrho(x)
	= \overline{\mathcal V}_2(d) - \overline{\mathcal V}_2(0)
	- x/h \bigl( \overline{\mathcal V}_1(d)- \overline{\mathcal V}_1(0) \bigr),
	\]
	where,  $\text{for }  \ell\in\{1,2\}$ and $s \in \{0,1,2\}$,
	\[
	\overline{\mathcal V}_{\ell}(d)
	=\frac{\mathcal V_{\ell}(h,d)}{\mathcal V_{2}(h,d)\mathcal V_{0}(h,d)-\mathcal V_{1}(h,d)^2},
	\qquad
	\qquad \mathcal V_s(h,d):= \sum_{i:\,X_i\ge d} K_h^+(X_i)\,(X_i/h)^s.
	\]
	Note that $A(s)$ is not a smooth function of $s$, but a step function that changes only when $s$ crosses a design point $X_i$ with $X_i \in (d,h)$. In particular, as $s$ increases past such a point $X_i$, the increment in $A(s)$ is
	\[
	\Delta A(X_i)
	= K_h^+(X_i)\,\varrho(X_i).
	\]
	We show below that
	\begin{equation}
		\text{(i)}\quad  \overline{\mathcal V}_1(0) \leq \overline{\mathcal V}_1(d),
		\qquad
		\text{(ii)}\quad \overline{\mathcal V}_2(0) \leq  \overline{\mathcal V}_2(d) , \label{equ:kappa_conditions}
	\end{equation}
	and thus the function \( \varrho(x) \) is linear in \( x \), and $\varrho(x)$ changes sign at most once on $(d,h)$, from positive to negative, as $x$ increases. Since $K_h^+(x) \ge 0$, it follows that the increments $\Delta A(X_i)$ also change sign at most once, from  positive to negative, as the ordered design points $X_i$ increase. Consequently, $A(s)$ is unimodal as a function of $s$ and $A(h) =T_2$. It follows that
	$ \frac{\partial}{\partial s} v_\Delta(s,h,d)$ changes its sign at most once on $s\in (d,h)$.	As it is also clear that  $ \frac{\partial}{\partial s} v_\Delta(d,h,d)\leq0 $ and $ \lim{s \uparrow h} \frac{\partial}{\partial s} v_\Delta(s,h,d)=0 $, it follows that $v_\Delta(s,h,d)\leq 0$ for all $s>0$.

	\medskip
	We now verify conditions (i) and (ii).
	Let \( 0 = d_0 < d_1 < \dots < d_J = d \) be a partition of \( [0,d] \) such that  one distinct realization of the running variable
	\( X_i \) lies in each half-open interval \( [d_{j-1}, d_j) \).
	Then
	\[
	\overline{\mathcal V}_{\ell}(0) - \overline{\mathcal V}_{\ell}(d)
	= \sum_{j=1}^J
	\bigl(
	\overline{\mathcal V}_{\ell}(d_{j-1}) - \overline{\mathcal V}_{\ell}(d_j)
	\bigr),
	\qquad \ell \in \{1,2\}.
	\]
	It therefore suffices to show that 	$\overline{\mathcal V}_{\ell}(d_{j-1})  - \overline{\mathcal V}_{\ell}(d_j)\leq 0$ for all $j$.
	We note that
	\[
	\overline{\mathcal V}_{\ell}(d_{j-1})-\overline{\mathcal V}_{\ell}(d_j)
	=\frac{\mathcal A_\ell}{D(d_{j-1})D(d_j)},
	\]
	where
	\[
	D(d):=\mathcal V_2(h,d)\mathcal V_0(h,d)-\mathcal V_1(h,d)^2 >0
	\]
	and
	\[
	\mathcal A_\ell
	:=\mathcal V_\ell(h,d_{j-1})D(d_j)-\mathcal V_\ell(h,d_j)D(d_{j-1}).
	\]
	The sign of the difference is determined by $\mathcal A_\ell$ since $D(d_{j-1}),D(d_j)>0$.

	Fix $j$ and let $x$ denote the (unique) value of the running variable in $[d_{j-1},d_j)$ (unique by construction), and let
	$m$ be the number of observations with $X_i=x$ (so $m>1$ only if $X$ has mass points). Since moving the cutoff from $d_{j-1}$ to $d_j$ removes exactly the observations with $X_i=x$,
	we have the one-step update formulas
	\begin{equation}\label{eq:one_step_update}
		\mathcal V_s(h,d_{j})=\mathcal V_s(h,d_{j-1})-\delta \,x^s/h^s,
		\qquad s=0,1,2,
	\end{equation}
	where $\delta =mK_h^+(x)$.
	Using \eqref{eq:one_step_update}, we have
	\begin{align*}
		D(d_j)
		&=(\mathcal V_2-\delta (x/h)^2)(\mathcal V_0-\delta)-(\mathcal V_1-\delta x/h)^2\\
		&=(\mathcal V_2\mathcal V_0-\delta (x/h)^2\mathcal V_0)- \delta(\mathcal V_2-\delta (x/h)^2)-(\mathcal V_1^2-2 \mathcal V_1 \delta x/h+ (\delta x/h)^2)\\
		&=(\mathcal V_2\mathcal V_0-\mathcal V_1^2)  - \delta ((x/h)^2\mathcal V_0 + \mathcal V_2-\delta (x/h)^2 - 2 \mathcal V_1 x/h + \delta (x/h)^2)\\
		&=D(d_{j-1})-\delta\bigl((x/h)^2\mathcal V_0+\mathcal V_2-2(x/h)\mathcal V_1\bigr),
	\end{align*}
	where all $\mathcal V_s$ on the right-hand side are evaluated at $(h,d_{j-1})$.
	Substituting these identities into the definition of $\mathcal A_\ell$ yields
	\begin{align*}
		\mathcal A_\ell
		&=\mathcal V_\ell\Bigl(D-\delta((x/h)^2\mathcal V_0+\mathcal V_2- 2x/h\mathcal V_1)\Bigr)
		-(\mathcal V_\ell - \delta (x/h)^\ell)D\\
		&=\mathcal V_\ell\Bigl(-\delta((x/h)^2\mathcal V_0+\mathcal V_2- 2x/h\mathcal V_1)\Bigr)
		+ \delta (x/h)^\ell (\mathcal V_2\mathcal V_0-\mathcal V_1^2)
	\end{align*}
	again with all $\mathcal V_s$ and $D$ evaluated at $(h,d_{j-1})$.
	For \( \ell = 1 \), this term simplifies to
	\begin{align*}
		\mathcal A_1
		&=\mathcal V_1\Bigl(-\delta((x/h)^2\mathcal V_0+\mathcal V_2- 2(x/h)\mathcal V_1)\Bigr)
		+ \delta (x/h) (\mathcal V_2\mathcal V_0-\mathcal V_1^2)\\
		&=\delta (-(x/h)^2\mathcal V_0\mathcal V_1 - \mathcal V_1 \mathcal V_2+ (x/h)\mathcal V_1^2
		+  (x/h) \mathcal V_2\mathcal V_0)\\
		&=\delta ( \mathcal V_1 (x/h\mathcal V_1 -\mathcal V_2)
		+  x/h \mathcal V_0 (\mathcal V_2 - x/h\mathcal V_1  )) \\
		&=\delta ( \mathcal V_1 - x/h \mathcal V_0)  (x/h\mathcal V_1 -\mathcal V_2) \leq 0
	\end{align*}
	
	where the last inequality follows because every observation contributing
	to the moments evaluated at $(h,d_{j-1})$ satisfies $X_l\geq x$. Hence,
	for \(s\in\{0,1\}\),
	\[
	\mathcal V_s(h,d_{j-1})\frac{x}{h}
	=
	\sum_{l:\,X_l\geq d_{j-1}}
	K_h^+(X_l)
	\left(\frac{X_l}{h}\right)^s
	\frac{x}{h}
	\leq 	
	\sum_{l:\,X_l\geq d_{j-1}}
	K_h^+(X_l)
	\left(\frac{X_l}{h}\right)^{s+1}
	=
	\mathcal V_{s+1}(h,d_{j-1}),
	\]
	where the inequality uses \(x/h\leq X_l/h\) for every contributing
	observation. Since the kernel weights are nonnegative, it follows that
	\(\mathcal A_1\leq0\).

	For \( \ell = 2 \), it holds that
	\begin{align*}
		\mathcal A_2
		=\mathcal V_2\Bigl(-\delta(\left(\frac{x}{h}\right)^2\mathcal V_0+\mathcal V_2- 2\frac{x}{h}\mathcal V_1)\Bigr)
		+ \delta \left(\frac{x}{h}\right)^2 (\mathcal V_2\mathcal V_0-\mathcal V_1^2)
		= - \delta( \mathcal V_2-  \frac{x}{h} \mathcal V_1)^2\leq 0
	\end{align*}
	Combining the two cases, we conclude that equation~\eqref{equ:kappa_conditions} holds, which completes this proof.

\subsection{Proof of Theorem~\ref{theorme::local_alternatives}}

	We prove the theorem in two parts.

	\noindent\textbf{Part 1.} 	We first consider the test statistic $t_\Delta(h,d)$. The following decomposition holds under the local alternative $H_{1,n}$:
	\begin{align*}
		t_\Delta(h,d)
		&= \frac{\sum_{i \in [n]} w_{i,1,\Delta}(h, d) \mu(X_i)}{\wh s_{\Delta}(h,d)}
		+ \frac{\sum_{i \in [n]} w_{i,1,\Delta}(h, d) \eta_n(X_i)}{\wh s_{\Delta}(h,d)}
		+ \frac{\sum_{i \in [n]} w_{i,1,\Delta}(h, d) \varepsilon_i}{\wh s_{\Delta}(h,d)} \\
		&\equiv T_{\Delta,1}(h,d) + T_{\Delta,2}(h,d) + Z_\Delta(h,d).
	\end{align*}
	In the following, we study each term separately. By the arguments of Lemma~\ref{lemma::clt}, the term $Z_\Delta(h,d)$ converges in distribution to $N(0,1)$. To study  $T_{\Delta,1}$, we use known bias expansions and scaling relations, and conclude that
	\begin{align*}
		T_{\Delta,1}(h,d)
		&	= - \frac{M h^2 (B_{K}(c) -B_{K}(0))}{\sqrt{\frac{1}{nh}} \sigma_\tau \sqrt{S_{K,\Delta}(c)}} + o_P(1)	=  -\frac{M h_0^{5/2} (B_{K}(c) -B_{K}(0))}{\sigma_\tau \sqrt{S_{K,\Delta}(c)}} + o_P(1),
	\end{align*}
	where we used that $h=h_0 \cdot n^{-1/5}$. To study $T_{\Delta,2}(h,d)$, note that under the local alternative $H_{1,n}$ we have that
	$$ \eta_n(x)= \sgn(x)\frac{ \lambda}{d^{2}\sqrt{nh}} \,\frac{(x-\sgn(x)d)^2}{2} \1{|x|<d}.$$
	It then holds that
	\begin{align*}
		T_{\Delta,2}(h,d)		& =	\frac{-\sum_{i \in [n]} w_{i,1}(h, 0) \eta_n(X_i) \1{|X_i|<d}}{\wh s_\Delta(h,d)} \end{align*}
	where we used that  $\sum_{i \in [n]} w_{i,1}(h, d) \eta_n(X_i) \1{|X_i|<d}=0$. It further follows that
	\begin{align*}
		T_{\Delta,2}(h,d)		&	=	-\frac{\sum_{i \in [n]} w_{i,1}(h, 0) \sgn(X_i) (\lambda / (d^2 \sqrt{nh}) (X_i-\sgn(X_i)d)^2 /2)  \1{|X_i|<d}}{\wh s_\Delta(h,d)} \\
		&=	-\frac{\lambda  }{d^2\sqrt{nh}} \frac{\sum_{i \in [n]} w_{i,1}(h,0)  (X_i-d)^2  \1{0 \leq  X_i<d}}{s_\Delta(h,d)} + o_P(1)
	\end{align*}
	where we used the symmetry of the expression and consistency of the standard error.
		Substituting the form of the standard deviation, we get that
	\begin{align*}
		T_{\Delta,2}(h,d)
		&=	-\lambda \frac{ 1 }{c^2 h^2} \frac{\sum_{i \in [n]} w_{i,1}(h,0)  (X_i-d)^2  \1{0 \leq  X_i<d}}{ \sigma_\tau \sqrt{S_{K,\Delta}(c)} } + o_P(1)
	\end{align*}
	where we used that $c = d/h$.
	Using standard kernel calculations, it then follows that
	\begin{align*}
		T_{\Delta,2}(h,d)
		&=	-\lambda \frac{ 1 }{c^2 } \frac{ \int_0^c J_K(u,0) (u - c)^2 K(u) \, du}{ \sigma_\tau \sqrt{S_{ K,\Delta}(c)} } + o_P(1)
	\end{align*}
	Combining both leading terms,  it follows that
\[
T_{\Delta,1}(h,d)+T_{\Delta,2}(h,d)
=
\mathcal D_\Delta(c,h_0,K)
-\lambda\mathcal L_\Delta(c,K)
+o_P(1).
\]
The result follows by  Slutsky's theorem.
	
	\bigskip
	\noindent\textbf{Part 2.}
	For the second part of the Theorem,  we have that
	\begin{align*} t_\Gamma(h,d) & = 	\frac{\sum_{i \in [n]} w_{i,1, \Gamma}(h, d) \mu(X_i)}{\wh s_{\Gamma}(h,d)} -  \frac{\sum_{i \in [n]} w_{i,1}(d, 0) \eta_n(X_i)}{\wh s_{\Gamma}(h,d)} + \frac{\sum_{i \in [n]} w_{i,1, \Gamma}(h, d) \varepsilon_i}{\wh s_{\Gamma}(h,d)}
		\\
		& \equiv T_{\Gamma,1}(h,d) +T_{\Gamma,2}(h,d) + Z_\Gamma(h,d).
	\end{align*}
	It follows from Lemma~\ref{lemma::clt} that $Z_\Gamma(h,d)$ is asymptotically standard normal. To study $T_{\Gamma,1}(h,d)$, one can use similar arguments as above to show that
	\begin{align*}
		T_{\Gamma,1}(h,d)
		&	=  - \frac{M h_0^{5/2} (B_{K}(c) - c^2 B_{K}(0))}{\sigma_\tau \sqrt{S_{K, \Gamma}(c)}} + o_P(1),
	\end{align*}
	where we used that $h=h_0 \cdot n^{-1/5}$. To study $T_{\Gamma,2}(h,d)$, we then note that
	under the local alternative $H_{1,n}$, it holds again that
	$$ \sum_{i \in [n]} w_{i,1}(h, d) \eta_n(X_i) \1{|X_i|<d}=0$$ and further that
	\begin{align*}
		T_{\Gamma,2}(h,d)&= -\frac{\sum_{i \in [n]} w_{i,1}(d, 0) \eta_n(X_i)}{s_{\Gamma}(h,d)} + o_P(1)
	\end{align*}
	Using the properties of local linear weights, and that $h=h_0 \cdot n^{-1/5}$, it follows that
	\begin{align*}
		T_{\Gamma, 2}(h,d)& = -\frac{ \frac{\lambda}{\sqrt{nh}} +  \frac{\lambda}{2 d^2\sqrt{nh}}  \sum_{i \in [n]} w_{i,1}(d,0) \text{sign}(X_i) X_i^2}{s_{\Gamma}(h,d)} + o_P(1)\\
		& = -\frac{\lambda}{\sqrt{nh}} \frac{ 1 +   B_K(0)}{s_{\Gamma}(h,d)} + o_P(1)\\
		& = -\lambda\frac{ 1 +   B_K(0)}{\sigma_{\tau} \sqrt{S_{K, \Gamma}(c)}} + o_P(1)
	\end{align*}
		Combining both leading terms,  it follows that
	\[
	T_{\Gamma,1}(h,d)+T_{\Gamma,2}(h,d)
	=
	\mathcal D_\Gamma(c,h_0,K)
	-\lambda\mathcal L_\Gamma(c,K)
	+o_P(1).
	\]
	The result follows by  Slutsky's theorem.
	\qed

\subsection{Proof of Proposition~\ref{prop:pointest_smalldonut_vp}}
	We note that the donut RD estimator is equivalent to a standard RD estimator with a new kernel function of the form $K_c(u)=K(u)\1{|u|>c}$. The result follows by applying standard local polynomial arguments and by noting that the kernel constant changes as claimed. \qed

\subsection{Proof of Proposition~\ref{prop:inference_smalldonut}}
	Proposition~\ref{prop:inference_smalldonut} directly  follows from  Proposition~\ref{prop:pointest_smalldonut_vp} by noting that under the given bandwidth the bias term is asymptotically negligible relative to the standard deviation. \qed

\section{Additional Details on Fuzzy RD Designs}\label{appendix:fuzzy}

In this appendix, we explain how the analysis from the main body of the paper can be extended to fuzzy donut regression discontinuity  designs, in which the treatment probability jumps at the cutoff, but not necessarily from zero to one. We describe how to implement the respective procedures and sketch the theoretical rationale behind them, but do not provide full proofs (these follow standard arguments). We use notation that is analogous to that in the main body of the paper as much as possible.  For
a generic random variable $V_i$, write
\[
        \mu_V(x)=\E(V_i\mid X_i=x),
        \qquad
        \tau_V=\mu_{V+}-\mu_{V-},
\]
where $\mu_{V+}$ and $\mu_{V-}$ denote the right and left limits at the cutoff.  In a
fuzzy RD design the conventional estimand is
\[
        \theta=\frac{\tau_Y}{\tau_T},
\]
provided that $\tau_T\neq0$.  The local linear fuzzy donut RD estimator is the Wald
ratio
\[
        \wh\theta(h,d)=\frac{\wh\tau_Y(h,d)}{\wh\tau_T(h,d)},\qquad
        \wh\tau_V(h,d)=\sum_{i \in [n]} w_i(h,d)V_i,
        \qquad V\in\{Y,T\},
\]
where the weights $w_i(h,d)$ are the sharp donut RD weights defined in
Section~\ref{sec::setup}. As in the sharp case, we postulate the existence of functions $\mu_Y^*$ and $\mu_T^*$ that
agree with the observed conditional expectation functions outside the donut but not
necessarily inside it:
\[
        \mu_V^*(x)=\mu_V(x)\quad\textnormal{for } |x|\ge d,
        \qquad V\in\{Y,T\}.
\]
We define the jumps at the cutoff and the fuzzy donut RD target by
\[
        \tau_V^*=\mu^*_{V+}-\mu^*_{V-},
        \qquad
        \theta^*=\frac{\tau_Y^*}{\tau_T^*},
\]
whenever $\tau_T^*\neq0$.  Curvature restrictions are expressed in terms of the
same function class used in the sharp RD analysis.  Recall that $\mathcal F(M)$ is the
class of functions with bounded second derivative on either side of the cutoff, as
defined in Section~\ref{sec::setup}.  For any $\eta\ge0$, define
\[
        \mathcal F_\eta(M)
        =
        \{m\in\mathcal F(M): |m_+(0)-m_-(0)|\ge\eta\}.
\]
We maintain that
\begin{equation}\label{eq:newB2_fuzzy_class}
        (\mu_Y^*,\mu_T^*)
        \in
        \mathcal F(M_Y)\times\mathcal F_\eta(M_T).
\end{equation}
This condition applies the sharp RD curvature restriction to the reduced form and the first stage and adds
the lower-bound parameter $\eta$ to the first-stage  function.  The case
$\eta>0$ corresponds to strong identification.  The case $\eta=0$ imposes no lower
bound on the first-stage jump because $\mathcal F_0(M_T)=\mathcal F(M_T)$.

There are then, broadly speaking, two useful regimes under which to analyze fuzzy donut RD designs.  The first regime imposes a strong first stage and uses small-donut asymptotics.  In that case the Wald ratio
admits a linear expansion around the conventional estimand $\theta$, and the problem reduces to a sharp donut RD problem for a
suitable auxiliary outcome.  The second regime allows weak first stages and permits either
small- or large-donut asymptotics.  In that case the ratio estimator is not consistent and cannot be used as a basis for inference on $\theta^*$, but Anderson--Rubin--type confidence sets as
in \citet{noack2024bias} and componentwise specification tests remain available.

\subsection{Strong First Stage and Small-Donut Asymptotics}

We first consider the case in which the donut is small relative to the bandwidth, as in
Assumption~\ref{ass:sd}, and the first stage is strong, i.e.,  $\eta>0$.  
The analysis in this subsection is based on the insight that if the treatment discontinuity is bounded away from zero and standard regularity conditions hold, then under small-donut asymptotics the fuzzy donut RD estimator is asymptotically equivalent to a sharp donut RD estimator based on a modified outcome:
\[
\widehat{\theta}(h,d) - \theta
=
\sum_{i \in [n]} w_i(h,d) U_i
+ o_P(h^2+1/\sqrt{nh}) \quad \text{ with } \quad U_i = \frac{Y_i - \tau_Y}{\tau_T}
- \frac{\tau_Y}{\tau_T^2}(T_i - \tau_T).
\]
For further analysis, it will be useful to define the functions
\[
     \mu_U(x)=\frac{\mu_Y(x)-\theta\mu_T(x)}{\tau_T}, \quad   \mu_U^*(x)=\frac{\mu_Y^*(x)-\theta\mu_T^*(x)}{\tau_T}.
\]
Note that $\tau_U=0$ by construction. Moreover, under the specification
null, $\tau_U^*=0$. We have that
\[
        \mu_U^*\in\mathcal F(M_U),
        \qquad
        M_U=\frac{M_Y+|\theta|M_T}{|\tau_T|}.
\]

\subsubsection{Point Estimation}
After linearization, the same kernel constants that govern sharp donut RD determine the
leading bias and variance of the fuzzy donut estimator around $\theta$. In particular, if
$\mu_V^*(x)=\mu_V(x)$ for all $|x|<d$ and $V\in\{Y,T\}$, and standard regularity conditions hold, then
\[
        \frac{\wh\theta(h,d)-\theta-b_\theta(h,d)}{s_\theta(h,d)}
        \stackrel{d}{\to}N(0,1),
\]
where
\[
        b_\theta(h,d)
        =
        h^2B_K(c)
        \frac{(\mu_{Y+}^{\prime\prime}-\mu_{Y-}^{\prime\prime})
        -\theta(\mu_{T+}^{\prime\prime}-\mu_{T-}^{\prime\prime})}
        {2\tau_T}
        +o_P(h^2),
\]
and
\[
        s_\theta^2(h,d)
        =
        \frac{1}{nh}S_K(c)
        \left(
        \frac{\sigma_{U+}^2}{f_+}
        +
        \frac{\sigma_{U-}^2}{f_-}
        \right)
        +o_P\left(\frac{1}{nh}\right).
\]
Here $\sigma_{U\pm}^2=\lim_{x\to0^\pm}\Var(U_i\mid X_i=x)$.

\subsubsection{Bias-Aware Confidence Intervals}

The point-estimation result suggests using the same bias-aware logic as in the sharp
case, but applied to the linearized fuzzy estimator.  Let
\[
        s_\theta^2(h,d)
        =
        \Var\left(\sum_{i \in [n]} w_i(h,d)U_i\mid X_1,\ldots,X_n\right),
\]
and let $\wh s_\theta^2(h,d)$ be a consistent feasible estimator, for example a
nearest-neighbor variance estimator applied to a generated outcome
$\wh U_i=(Y_i-\wh\theta T_i)/\wh\tau_T$ formed from preliminary consistent estimates.
Define the worst-case bias
\[
        \overline b_\theta(h,d)
        =
        -\frac{ M_U}{2}
        \sum_{i \in [n]} w_i(h,d)X_i^2\sgn(X_i).
\]
Since $M_U$ depends on the unknown quantities $\theta$ and $\tau_T$, creating a feasible procedure requires either bounding these quantities  or replacing them with appropriate point estimates. We follow the second approach and define the ``feasible'' worst-case bias approximation
\[
        \wh{ \overline b}_\theta(h,d)
        =
        -\frac{ \wh M_U(h,d)}{2}
        \sum_{i \in [n]} w_i(h,d)X_i^2\sgn(X_i), \qquad \wh M_U(h,d) =\frac{M_Y+|\wh\theta(h,d)|M_T}{|\wh\tau_T(h,d)|}.
\]
We can then define a feasible bias-aware confidence interval for $\theta$ for this regime as
\[
        C_{\theta,n}(d)
        =
        \left[
        \wh\theta(h,d)
        \pm
        \textnormal{cv}_{1-\alpha}
        \left(
        \frac{\wh{\overline b}_\theta(h,d)}{\wh s_\theta(h,d)}
        \right)
        \wh s_\theta(h,d)
        \right],
\]
where $\textnormal{cv}_{1-\alpha}(r)$ denotes the $1-\alpha$ quantile of $|N(r,1)|$.
One can also construct robust bias-corrected fuzzy donut confidence intervals by
applying the usual higher-order local polynomial bias correction to the auxiliary
outcome $U_i$, with unknown quantities replaced by consistent estimates.

\subsubsection{Specification Testing} We proceed analogously for our two specification tests. 
We formalize the goal of testing for correct specification as the null hypothesis:
\[
        H_0:
        \mu_Y^*(x)=\mu_Y(x)\textnormal{ and }\mu_T^*(x)=\mu_T(x)
        \quad\textnormal{for all } |x|<d
\]
given our maintained assumptions that $\mu_Y^\star(x) = \mu_Y(x)$ and $\mu_T^\star(x) = \mu_T(x)$ for all $|x| \geq d$. Under standard regularity conditions, it can then be shown that
\begin{align*}
	\widehat \theta(h,d)-\widehat \theta(h,0)
	&=  \widehat\Delta_{\theta}(h, d)   +  o_P(h^2+1/\sqrt{nh})
	\text{ with }  \widehat\Delta_{\theta}(h, d)  = \sum_{i \in [n]} (w_i(h, d)-w_i(h, 0))U_i,\\
	\widehat \theta(h,d)-\widehat \theta(d,0)
	&=  \widehat\Gamma_{\theta}(h, d)   +  o_P(h^2+1/\sqrt{nh}) \text{ with }  \widehat\Gamma_{\theta}(h, d)  = \sum_{i \in [n]} (w_i(h, d)-w_i(d, 0))U_i,
\end{align*}
with $U_i$ defined as above. Hence, inference can proceed analogously to Section~\ref{sec::specification_testing} by noting that the  differences between the two fuzzy donut RD estimates are again well-approximated by quantities that behave like sharp RD objects. Specifically,
let 
\begin{align*}
        s_{\Delta,\theta}^2(h,d)
        =
        \Var\left(\sum_{i \in [n]} (w_i(h, d)-w_i(h, 0))U_i\mid X_1,\ldots,X_n\right),\\
         s_{\Gamma,\theta}^2(h,d)
        =
        \Var\left(\sum_{i \in [n]} (w_i(h, d)-w_i(d, 0))U_i\mid X_1,\ldots,X_n\right),
\end{align*}
and let $\wh s_{\Delta,\theta}^2(h,d)$ and $\wh s_{\Gamma,\theta}^2(h,d)$ be  consistent feasible estimators, for example a
nearest-neighbor variance estimator applied to a generated outcome
$\wh U_i=(Y_i-\wh\theta T_i)/\wh\tau_T$ formed from preliminary consistent estimates.
Also define the feasible worst-case bias approximations
\begin{align*}
        \wh {\overline b}_{\Delta,\theta}(h,d)
        &=
        -\frac{ \wh M_U}{2}
        \sum_{i \in [n]} (w_i(h, d)-w_i(h, 0))X_i^2\sgn(X_i), \\\wh{ \overline b}_{\Gamma,\theta}(h,d)
        &=
        -\frac{ \wh M_U}{2}
        \sum_{i \in [n]} (w_i(h, d)-w_i(d, 0))X_i^2\sgn(X_i),
\end{align*}
with $\wh M_U$ as defined above. The test statistics then are 
\[
        t_{\Delta,\theta}
        =
        \frac{\widehat \theta(h,d)-\widehat \theta(h,0)}{\wh s_{\Delta,\theta}(h,d)} \text{ and } t_{\Gamma,\theta}
        =
        \frac{\widehat \theta(h,d)-\widehat \theta(d,0)}{\wh s_{\Gamma,\theta}(h,d)}.
\]

Each of the bias-aware specification tests rejects $H_0$ with nominal level $\alpha$ if the corresponding condition holds
\[
        |t_{\Delta,\theta}|
        >
       \textnormal{cv}_{1-\alpha}
        \left(
        \frac{\wh{\overline b}_{\Delta,\theta}(h,d)}{\wh s_{\Delta,\theta}(h,d)}
        \right)\text{ and  }  |t_{\Gamma,\theta}|
        >
       \textnormal{cv}_{1-\alpha}
        \left(
        \frac{\wh{\overline b}_{\Gamma,\theta}(h,d)}{\wh s_{\Gamma,\theta}(h,d)}
        \right).
\]

\subsection{Possibly Weak First Stage} We now consider the case in which the first stage is possibly weak, in the sense that~\eqref{eq:newB2_fuzzy_class} only holds with $\eta=0$. In this case, the linearization argument used in the previous subsection breaks down and thus cannot be used.

\subsubsection{Point Estimation}

Under the  restriction that $(\mu_Y^\star, \mu_T^\star) \in \mathcal{F}(M_Y) \times \mathcal{F}(M_T)$,
the fuzzy RD parameter of interest is, in general, only partially identified. It is easy to see that uniformly consistent point estimation is not possible in such cases under either small- or large-donut asymptotics. We therefore do not further investigate the properties of the fuzzy RD estimator. Following arguments analogous to those in Remark~\ref{remark::largedonut} in the main body of the paper, the identified set can be shown to be 
\[
\left\{ \vartheta \in \mathbb{R} \; :\;
0 = \tilde{\tau}_Y - \vartheta \, \tilde{\tau}_T 
\text{ for some }
(\tilde{\tau}_Y, \tilde{\tau}_T) \in [\tau_\textnormal{Y,Lin}(d) \pm M_Y d^2] \times [\tau_\textnormal{T,Lin}(d) \pm M_T d^2]
\right\},
\]
$$\text{where } 
\tau_\textnormal{V,Lin}( d)= \mu_V(d)- \mu_V(-d) - d(\mu_V'(d) +\mu_V'(-d)) \text{ for } V \in \{T,Y\}.$$

This set is either a closed interval, the full real line, a half-line, or the union of two half-lines.

\subsubsection{Bias-Aware Confidence Intervals}

Despite a lack of point identification, one can construct valid Anderson-Rubin-type bias-aware confidence sets for the fuzzy donut RD estimand following the arguments of Section~\ref{sec::CIs} and \cite{noack2024bias}:
\begin{align}
	\mathcal{C}_{\textnormal{ar}}^{\alpha}(d)
	=\left\{\vartheta: |\widehat\tau_{Y}(h,d)- \vartheta \widehat\tau_{T}(h,d) | \leq \textnormal{cv}_{1-\alpha}\!\left(\frac{\overline{b}_{AR}(h,d,\vartheta)}{
	\widehat{s}_{AR}(h,d,\vartheta)}\right)\widehat s_{AR}(h,d,\vartheta) \right \},
\end{align}
where the maximal bias is given by $\overline{b}_{AR}(h,d,\vartheta)= - \frac{M_Y + |\vartheta|  M_T}{2}  \sum_{i \in [n]} w_{i}(h,d) X_i^2  \sgn(X_i)$ and $\widehat s_{AR}(h,d,\vartheta)^2$ is a consistent estimate of the variance of $\widehat\tau_{Y}(h,d)- \vartheta \widehat\tau_{T}(h,d)$, $s_{AR}(h,d,\vartheta)^2$, with
${ s_{AR}(h,d,\vartheta)^2= \sum_{i \in [n]} w_i(h,d)^2  \left(\sigma_{Y,i}^2 - 2\vartheta\sigma_{YT, i}  + \vartheta^2 \sigma_{T,i}^2 \right)}, \text{ where } \sigma_{YT, i} = \Cov(Y_i, T_i\mid X_i)$ and $\sigma_{V,i}^2:=\Var(V_i\mid X_i)$ for $V \in \{Y,T\}$.
\subsubsection{Specification Testing}

We consider again the null hypothesis that the underlying regression functions coincide inside and outside the donut region for both the outcome and the treatment variables:
\[
H_0: \mu_Y^\star(x) = \mu_Y(x) \;\; \text{and} \;\; \mu_T^\star(x) = \mu_T(x)
\quad \text{for all } |x| < d,
\]
given our maintained assumptions that $\mu_Y^\star(x) = \mu_Y(x)$ and $\mu_T^\star(x) = \mu_T(x)$ for all $|x| \geq d$ and that $(\mu_Y^\star, \mu_T^\star) \in \mathcal{F}(M_Y) \times \mathcal{F}(M_T)$. 

Given the irregular behavior of the fuzzy RD point estimator under possible weak identification, we adapt our testing strategy and test the two components of the null hypothesis separately. For the case of the ``conventional'' testing strategy, this involves comparing  donut RD estimates and conventional RD estimates of the reduced form and first stage parameters. Following the discussion in Section~\ref{sec::specification_testing}, and using its notation with appropriate subscripts for the outcome and treatment variables, we can construct two test statistics for $\Delta_T$ and $\Delta_Y$ separately and adjust for multiple hypothesis testing using a Bonferroni correction.
We denote the resulting test statistic by
\[
T_{\Delta, \theta} = \max \left\{
|t_{T,\Delta}| - \textnormal{cv}_{1-\alpha/2}\!\left(\frac{\bar{b}_{T,\Delta}(h, d)}{\widehat s_{T,\Delta}(h,d)} \right),\;
|t_{Y,\Delta}| - \textnormal{cv}_{1-\alpha/2}\!\left(\frac{\bar{b}_{Y,\Delta}(h, d)}{\widehat s_{Y,\Delta}(h,d)} \right)
\right\}.
\]
The test rejects $H_0$ with nominal level $\alpha$ if $T_{\Delta, \theta} > 0$. The power of the test depends on the asymptotic framework and on the nature of the (local) alternatives under consideration. A test that compares donut and ``within-donut'' RD estimates can be conducted analogously.

\section{Small Monte Carlo Simulations} \label{sec:simulation}

	In this section, we report the results of a simple simulation study. We generate the running variable as $X_i\sim U(-1,1)$ and the outcome as $Y_i = \mu_L(X_i) + \varepsilon_i$, where $\varepsilon_i\sim N(0,0.5)$ independent of $X_i$ and $\mu_L(x) = \sgn(x)x^2 -L\sgn(x)((x-0.1\times\sgn(x))^2 -0.1^2\times\sgn(x))\1{|x|<0.1}$ for $L\in\{0,10,\ldots,40\}.$ We define $\mu^*(x) = \mu_0(x)$ so that $\tau^*=0$. The case $L=0$ then corresponds to a setting where donut RD estimation would not be  necessary, whereas for larger values of $L$ we have settings in which the conditional expectation of the observed data differs more and more extremely from its hypothetical counterpart over the area $(-0.1,0.1)$.
	We then compute conventional and donut RD estimates, corresponding bias-aware confidence intervals, and our two specification tests  with a triangular kernel, $d=0.1$ and $M=2$.\footnote{Computations are carried out in {\tt R} with the package {\tt RDHonest}, using the latter's default setting for bandwidth selection.} We use $n=1,000$ as the sample size and set the number of replications to 10,000. Our results can be summarized as follows.

\subsection*{Point Estimation} Table~\ref{table:pointest} shows  the simulated bias, standard deviation and root mean squared error (RMSE) of the conventional and the donut RD estimator for the various values of $L$. Note that the data-driven bandwidth selector for the conventional RD estimator chooses an average value of $h=0.49$.  The results in Table~\ref{table:pointest} can thus be compared with theoretical predictions under small donut asymptotics with $c\approx 0.19$. Focusing first on the case $L=0$ we see that the ratios of bias and standard deviation (and thus RMSE) between donut and conventional RD estimators are as predicted by theory. The properties of the donut RD estimator are then unaffected by $L$ as expected, whereas the bias, but not the standard deviation, of the conventional RD estimator increases with $L$. Note however only with the extreme value $L=40$ the RMSE of the donut estimator becomes smaller than that of the conventional one. This illustrates that extreme deviations from the usual assumptions are necessary for donut RD estimation to become the dominant point estimator.

\subsection*{Confidence Intervals} Table~\ref{table:ci} shows the empirical coverage rates and average lengths of the conventional and donut RD confidence intervals for the various values of $L$. Donut CIs have correct coverage for all values of $L$ as expected, and their average length remains the same across scenarios as well. Conventional CIs have correct coverage for $L=0$, which then deteriorates for larger values of $L$. The ratio of average lengths of the two CI types for $L=0$ is as predicted by our small donut asymptotic theory.

\subsection*{Specification Testing} Table~\ref{table:spec} shows the empirical rejection rates of the two specification tests we considered for the various values of $L$ and the usual nominal level $\alpha=0.05$. Both tests are seen to have correct size if $L=0$ and thus the null hypothesis holds. Both tests also exhibit increasing rejection rates in $L$ as expected. However, the alternative test based on $\widehat\Gamma$ exhibits strictly greater power than the conventional one based on $\widehat\Delta$ under the type of conditional expectation functions considered here.

	\begin{table}[!t]
	\centering
	\caption{Simulation Results: Point Estimation}\label{table:pointest}
	\begin{tabular}{@{} ccccccc @{}}
		\toprule
		& \multicolumn{2}{c}{Bias} & \multicolumn{2}{c}{Std. Dev.}& \multicolumn{2}{c}{RMSE}   \\
		$L$& Regular & Donut & Regular & Donut & Regular & Donut  \\
		\midrule
		0 & 0.043 & 0.115 & 0.099 & 0.162 & 0.108 & 0.198 \\
		10 & -0.026 & 0.115 & 0.099 & 0.161 & 0.102 & 0.198 \\
		20 & -0.094 & 0.115 & 0.100 & 0.162 & 0.137 & 0.199 \\
		30 & -0.161 & 0.117 & 0.101 & 0.162 & 0.190 & 0.200 \\
		40 & -0.229 & 0.116 & 0.102 & 0.162 & 0.251 & 0.199 \\
		\bottomrule
	\end{tabular}
	\bigskip
	\caption{Simulation Results: Confidence Intervals}\label{table:ci}
	\begin{tabular}{@{} ccccc @{}}
		\toprule
		& \multicolumn{2}{c}{CI Coverage} & \multicolumn{2}{c}{CI Length}  \\
		$L$& Regular & Donut & Regular & Donut \\
		\midrule
		0 & 0.954 & 0.948 & 0.430 & 0.764 \\
		10 & 0.962 & 0.949 & 0.431 & 0.764 \\
		20 & 0.884 & 0.946 & 0.430 & 0.764 \\
		30 & 0.699 & 0.946 & 0.430 & 0.763 \\
		40 & 0.445 & 0.947 & 0.430 & 0.763 \\
		\bottomrule
	\end{tabular}
	\bigskip
	\caption{Simulation Results: Specification Testing}\label{table:spec}
	\begin{tabular}{@{} ccc @{}}
		\toprule
		& \multicolumn{2}{c}{Rejection Frequency} \\
		$L$& $\widehat\Delta$ & $\widehat\Gamma$ \\
		\midrule
		0 & 0.053 & 0.052 \\
		10 & 0.119 & 0.152 \\
		20 & 0.232 & 0.345 \\
		30 & 0.396 & 0.593 \\
		40 & 0.565 & 0.803 \\
		\bottomrule
	\end{tabular}
\end{table}

\singlespacing

\bibliography{bibl}

@article{dowd2021donuts,
	title={Donuts and distant CATEs: Derivative bounds for RD extrapolation},
	author={Dowd, Connor},
	journal={Available at SSRN 3641913},
	year={2021}
}

@article{bau2020teacher,
Author = {Bau, Natalie and Das, Jishnu},
Title = {Teacher Value Added in a Low-Income Country},
Journal = {American Economic Journal: Economic Policy},
Volume = {12},
Number = {1},
Year = {2020},
Month = {February},
Pages = {62 - 96},
DOI = {10.1257/pol.20170243},
URL = {https://www.aeaweb.org/articles?id=10.1257/pol.20170243}}

@article{noack2024bias,
	title={Bias-Aware Inference in Fuzzy Regression Discontinuity Designs},
	author={Noack, Claudia and Rothe, Christoph},
	journal={Econometrica},
	volume={92},
	number={3},
	pages={687--711},
	year={2024},
	publisher={Wiley Online Library}
}

@article{almond2010estimating,
  title={Estimating marginal returns to medical care: Evidence from at-risk newborns},
  author={Almond, Douglas and Doyle Jr, Joseph J and Kowalski, Amanda E and Williams, Heidi},
  journal={Quarterly Journal of Economics},
  volume={125},
  number={2},
  pages={591--634},
  year={2010},
  publisher={MIT Press}
}

@article{armstrong2018optimal,
	title={Optimal inference in a class of regression models},
	author={Armstrong, Timothy and Koles{\'a}r, Michal},
	journal={Econometrica},
	volume={86},
	number={2},
	pages={655--683},
	year={2018},
	publisher={Wiley Online Library}
}

@article{imbens2019optimized,
	title={Optimized regression discontinuity designs},
	author={Imbens, Guido and Wager, Stefan},
	journal={Review of Economics and Statistics},
	volume={101},
	number={2},
	pages={264--278},
	year={2019},
	publisher={MIT Press}
}

@article{calonico2014robust,
	title={Robust nonparametric confidence intervals for regression-discontinuity designs},
	author={Calonico, Sebastian and Cattaneo, Matias D and Titiunik, Rocio},
	journal={Econometrica},
	volume={82},
	number={6},
	pages={2295--2326},
	year={2014},
	publisher={Wiley Online Library}
}

@article{imbens2012optimal,
	title={Optimal bandwidth choice for the regression discontinuity estimator},
	author={Imbens, Guido and Kalyanaraman, Karthik},
	journal={Review of Economic Studies},
	volume={79},
	number={3},
	pages={933--959},
	year={2012},
	publisher={Oxford University Press}
}

@article{armstrong2018simple,
	title={Simple and honest confidence intervals in nonparametric regression},
	author={Armstrong, Timothy B and Koles{\'a}r, Michal},
	journal={Quantitative Economics},
	volume={11},
	number={1},
	pages={1--39},
	year={2020},
	publisher={Wiley Online Library}
}

@article{abadie2014inference,
	title={Inference for misspecified models with fixed regressors},
	author={Abadie, Alberto and Imbens, Guido W and Zheng, Fanyin},
	journal={Journal of the American Statistical Association},
	volume={109},
	number={508},
	pages={1601--1614},
	year={2014},
	publisher={Taylor \& Francis}
}

@article{abadie2006large,
	author = {Abadie, Alberto and Imbens, Guido W.},
	title = {Large Sample Properties of Matching Estimators for Average Treatment Effects},
	journal = {Econometrica},
	volume = {74},
	number = {1},
	pages = {235--267},
	year = {2006}
}

@article{barreca2011saving,
	title={Saving babies? Revisiting the effect of very low birth weight classification},
	author={Barreca, Alan I and Guldi, Melanie and Lindo, Jason M and Waddell, Glen R},
	journal={Quarterly Journal of Economics},
	volume={126},
	number={4},
	pages={2117--2123},
	year={2011},
	publisher={Oxford University Press}
}

@ARTICLE{dong2014alternative,
	title={Alternative assumptions to identify LATE in fuzzy regression discontinuity designs},
	author={Dong, Yingying},
	journal={Oxford Bulletin of Economics and Statistics},
	volume={80},
	number={5},
	pages={1020--1027},
	year={2018},
	publisher={Wiley Online Library}
}

@BOOK{fan1996local,
  title = {Local polynomial modelling and its applications},
  publisher = {Chapman \& Hall/CRC},
  year = {1996},
  author = {Fan, J. and Gijbels, I.}
}

@ARTICLE{hahn2001identification,
  author = {Hahn, Jinyong and Todd, Petra and Van der Klaauw, Wilbert},
  title = {Identification and Estimation of Treatment Effects with a Regression-Discontinuity
	Design},
  journal = {Econometrica},
  year = {2001},
  volume = {69},
  pages = {201--209},
  number = {1},
  publisher = {Wiley Online Library}
}

@ARTICLE{kolesar2018discrete,
  author = {Kolesár, Michal and Rothe, Christoph},
  title = {Inference in Regression Discontinuity Designs with a Discrete Running
	Variable},
  journal = {American Economic Review},
  pages = {2277–-2304},
    volume = {108},
  year = {2018}
}

\end{document}